\documentclass[pageno]{jpaper}

\newcommand{\asplossubmissionnumber}{667}

\usepackage[normalem]{ulem}

\usepackage{graphicx}
\usepackage{braket}
\usepackage{soul}
\usepackage{booktabs}
\usepackage{adjustbox}

\usepackage{bm}
\usepackage{cite}

\usepackage[linewidth=1pt]{mdframed}

\usepackage{authblk}

\usepackage{tikz}
\newcommand*\circled[1]{\tikz[baseline=(char.base)]{
            \node[shape=circle,draw,inner sep=0.5pt] (char) {#1};}}
\usepackage{mathtools}

\usepackage[font=bf,skip=0pt]{caption}

\usepackage{tabularx}

\newcommand{\rev}{\textcolor{black}{ }\textcolor{black}}

\newcommand\blfootnote[1]{%
  \begingroup
  \renewcommand\thefootnote{}\footnote{#1}%
  \addtocounter{footnote}{-1}%
  \endgroup
}

\begin{document}

\title{Better Than Worst-Case Decoding for Quantum Error Correction}

\author[1]{Gokul Subramanian Ravi*}
\author[1]{Jonathan M. Baker}
\author[2]{Arash Fayyazi}
\author[1]{Sophia Fuhui Lin}
\author[3]{Ali Javadi-Abhari}
\author[2]{Massoud Pedram}
\author[1,4]{Frederic T. Chong}

\affil[1]{University of Chicago}%
\affil[2]{University of Southern California}%
\affil[3]{IBM Quantum}%
\affil[4]{Super.tech (a division of ColdQuanta)}%

\date{}
\maketitle

\thispagestyle{empty}
\blfootnote{\noindent * Correspondence: gravi@uchicago.edu}

\begin{abstract}

The overheads of classical decoding for quantum error correction grow rapidly with the number of logical qubits and their correction code distance. Decoding at room temperature is bottle-necked by refrigerator I/O bandwidth while cryogenic on-chip decoding is limited by area/power/thermal budget. 

To overcome these overheads, we are motivated by the observation that in the common case (over 90\% of the time), error signatures are fairly trivial with high redundancy / sparsity, since the error correction codes are over-provisioned to be able to correct for uncommon worst-case complex scenarios (to ensure substantially low logical error rates). If suitably exploited, these trivial signatures can be decoded and corrected with insignificant overhead, thereby alleviating the bottlenecks described above, while still handling the worst-case complex signatures by state-of-the-art means.

Our proposal, targeting Surface Codes, consists of: 

\circled{1}\ A lightweight decoder for decoding and correcting trivial common-case errors, designed for the cryogenic domain. The decoder is implemented for SFQ logic.

\circled{2}\ A statistical confidence-based technique for off-chip decoding bandwidth allocation, to efficiently handle rare complex decodes that are not covered by the on-chip decoder. 

\circled{3}\ A method for stalling circuit execution, for the worst-case scenarios in which the provisioned off-chip bandwidth is insufficient to complete all requested off-chip decodes.

In all, our proposal enables 70-99+\% off-chip bandwidth elimination across a range of logical and physical error rates, without significantly sacrificing the accuracy of state-of-the-art off-chip decoding. 
By doing so, it achieves 10-10000x bandwidth reduction over prior off-chip bandwidth reduction techniques.
Furthermore, it achieves a 15-37x resource overhead reduction compared to prior on-chip-only decoding.
\end{abstract}

\section{Introduction}

\begin{figure}[t]
\centering
\includegraphics[width=0.98\columnwidth,trim={0cm 0.5cm 0cm 0cm},clip]{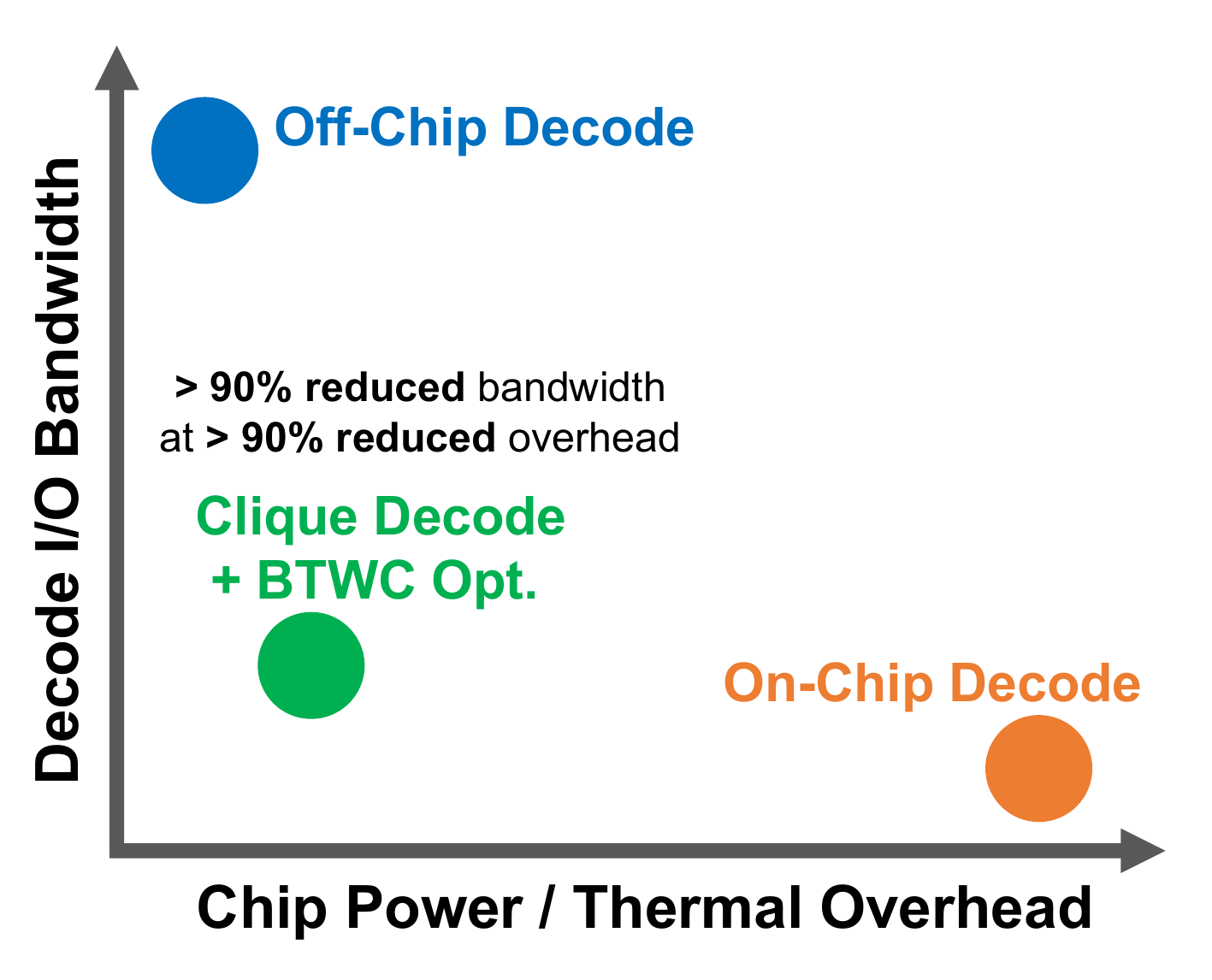}
\caption{
Off-chip QEC decoding requires high I/O bandwidth, whereas On-chip QEC can suffer from high thermal, power and area overheads. By adopting a Better Than Worst-Case design, our proposal achieves > 90\% bandwidth reduction while performing on-chip decoding at > 90\% reduction in on-chip overheads. 
}
\label{fig:clique_intro}
\end{figure}

\begin{figure*}[t]
\centering
\includegraphics[width=0.95\textwidth,trim={0cm 0cm 0cm 0cm},clip]{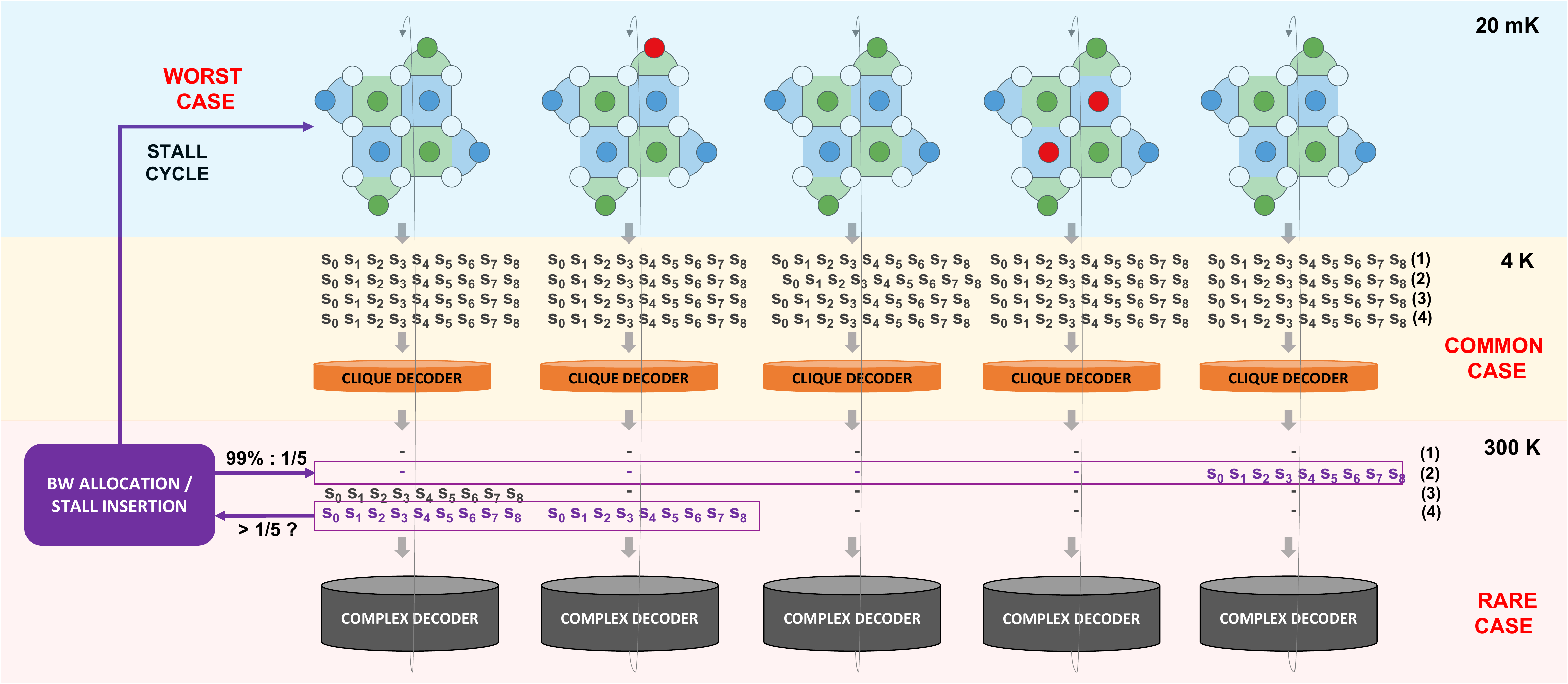}
\caption{
Illustrative overview of the proposal. 5 logical qubits with a code distance of 3 are shown. Every logical qubit generates error signatures (s0-s8) every cycle which need to be decoded and followed by the appropriate correction. We design a lightweight cryogenic on-chip Clique decoder which is able to accurately decode the common-case error signatures which are trivial to decipher. The rare complex scenarios are passed on to the complex decoder at room-temperature. Off-chip bandwidth is allocated appropriately so that there is sufficient bandwidth to handle all the off-chip decodes in most scenarios --- in this example, 99\% of the cycles requiring off-chip decodes have only 1 (out of a max 5) signatures to be decoded and bandwidth therefore provisioned as such. In the worst-case, if the number of off-chip decodes exceeds the provisioned bandwidth (when greater than 1 in this example), the quantum execution is stalled. Though new errors can occur in the stall cycle, the provisioned off-chip bandwidth is sufficient for all decodes to be resolved with minimal stall cycles.
}
\label{fig:clique_overview}
\end{figure*}
Quantum computation is a revolutionary information processing model that takes advantage of quantum mechanical phenomena. Quantum computers leverage superposition, interference, and entanglement, and this potentially gives them a significant computing advantage in solving intractable problems in domains of critical interest.

In today's Noisy Intermediate-Scale Quantum (NISQ)~\cite{preskill2018quantum} era of quantum computing, machines suffer from high error rates in the form of state preparation and measurement (SPAM) errors, gate errors, qubit decoherence, crosstalk, etc.
While in the near-term we will continue to target NISQ-favorable applications such as variational quantum algorithms~\cite{farhi2014quantum, peruzzo2014variational}, the long-term goal is to achieve quantum advantage on large-scale quantum algorithms like Shor's Factoring~\cite{Shor_1997} and Grover Search~\cite{Grover96afast}.
While quantum devices will continue to improve, qubit error rates,  even in the farther future, will be insufficient to directly run these large scale applications which demand high accuracy.
Therefore, they will require fault-tolerant systems~\cite{O_Gorman_2017} via quantum error correction (QEC), in which logical qubits are constructed from a collection of physical qubits in a way that the former performs many orders of magnitude better than the latter~\cite{Fowler_2012}.

The crux of QEC lies in qubit redundancy and classical support via error detection, decoding and correction.
Qubit redundancy, via mapping a large number of physical qubits to a single logical qubit, means that for physical qubits with reasonable error rates, the qubits with physical errors are fairly sparse in space and time.
When errors are sparse, the corresponding error signatures for each logical qubit are often \emph{trivial} (i.e., low Hamming weight) and can be accurately deciphered by a classical decoder with relative ease, far from utilizing its full decoding capabilities, and the appropriate corrections can be added to the erring qubits.
On the rare occasion when physical errors inopportunely congregate, the resulting \emph{complex} error signatures can be more difficult to decode (and sometimes impossible for a given error code specification), thus requiring the decoder to use its decoding capabilities to the fullest.
Accounting for these rare cases is still critical because incorrect error signature handling will result in logical errors that condemn the quantum computation, and likely the entire application.
Thus, to avoid the costly worst-case ramifications of failed application executions, it is expected that a) qubit redundancy will be set as high as possible so that complicated errors are rare, and b) decoders will be designed for highest accuracy, at correspondingly high resource cost, to near-perfectly decipher the even the rare complicated error signatures.
But designing for the extremely scarce worst-case, while clearly important (as discussed above), means that there is significant under-utilization of QEC resources and their capabilities in the more common trivial scenarios, thus presenting opportunity for optimizations and improvements.
Inspired by Better Than Worst-Case (BTWC)\footnote{Coined by reputed computer architect Bob Colwell of Pentium fame.} designs in the classical computing world~\cite{austin2005opportunities,cong2014better}, this work utilizes BTWC for quantum error correction to leverage the aforementioned opportunity. 

Now, we dive deeper into the resource constraints of QEC decoding, which is traditionally pursued via two alternate approaches.
The first, and more thoroughly studied, approach is off-chip decoding, in which the decoding is performed at room temperature via software ~\cite{Dennis_2002}, FPGAs~\cite{Lilliput} or custom hardware~\cite{AFS}.
Off-chip decoding, shown in blue in Fig.\ref{fig:clique_intro}, can require multiple Gbps of off-chip error data transmission ~\emph{per logical qubit}~\cite{AFS}, due to high coded qubit redundancy coupled with execution latency constraints~\cite{NISQ+}. 
Provisioning for such considerable bandwidth at the quantum-classical interface (i.e., between the cryogenic refrigerator and room temperature) is a serious scalability challenge due to limited I/O wiring.
The second approach, which has been more recently proposed, is on-chip decoding~\cite{NISQ+,QECOOL} at cryogenic temperatures. 
While this alleviates the I/O bandwidth constraint, cryogenic classical controllers located inside the refrigerator are subject to area, power and thermal dissipation constraints.
To meet these constraints, on-chip decoding is limited in its accuracy --- thus, achieving the target logical error rates for long-term quantum goals, via on-chip decoding alone, will be challenging.

Both these approaches suffer from the bottlenecks discussed above because they are provisioned to handle the rarely occurring / worst-case complex error signatures (albeit with different accuracy).
The fundamental insight in our proposal, inspired by BTWC philosophy, is to decouple the handling of the common trivial error signatures from the handling of the rare complex error signatures.
We propose a low overhead on-chip `Clique' decoder tailored to Surface Codes~\cite{Surface-Codes} that will identify and handle trivial errors, which are a high fraction of the total error signatures (over 90\% in most scenarios).
Surface Codes, which are among the most promising QEC codes due to their high error thresholds, are particularly amenable to trivial-case lightweight decoding due to their high locality~\cite{Fowler_2012}.
In addition to the Clique decoder, we propose optimizations that gracefully hand over the rare complex error signatures to any highly accurate state-of-the-art off-chip decoder (we use Minimum Weight Perfect Matching~\cite{Dennis_2002}), in a manner such that the provisioned off-chip transfer bandwidth is minimized while accounting for other execution constraints.
As shown in Fig.\ref{fig:clique_intro}, our proposal is uniquely positioned to mitigate both the I/O bandwidth bottleneck as well as the chip power/area/thermal bottleneck, and is thus a critical step towards a practical scalable future for quantum error correction.

In summary, we propose \textbf{\emph{Better Than Worst-Case (BTWC) Decoding for Quantum Error Correction}}, targeting Surface Codes, consisting of three components that are described below and illustratively summarized in Fig.~\ref{fig:clique_overview}:

\circled{1}\ \textbf{On-chip Clique Decoder:} An extremely lightweight decoder to detect, decode and correct the common-case isolated errors, designed for the cryogenic domain. We implement and evaluate the decoder for SFQ logic.

\circled{2}\ \textbf{Statistical Off-chip Bandwidth Allocation:} A statistical confidence-based technique for off-chip decoding bandwidth allocation, to efficiently handle the rare complex decodes that are not covered by the Clique Decoder. 

\circled{3}\ \textbf{Decode-Overflow Execution Stalling:} A method for stalling circuit execution, by means of idle gate insertion, for the worst-case scenarios in which the provisioned off-chip bandwidth is insufficient to complete all requested off-chip decodes.

\textbf{Key results and insights:}

\circled{1}\ In all, BTWC Decoding enables 70-99+\% off-chip bandwidth elimination across a range of logical and physical error rates, without significantly sacrificing the accuracy of state-of-the-art off-chip decoding. 

\circled{2}\ By doing so, it achieves 10-10000x bandwidth reduction over prior off-chip bandwidth reduction technique AFS~\cite{AFS}. 

\circled{3}\ Furthermore, it achieves a 15-37x resource overhead reduction compared to prior on-chip-only decoding NISQ+~\cite{NISQ+}.

\circled{4}\ Most importantly, we showcase that BTWC design is a critical step towards a practical scalable future for quantum error correction. While we focus on decoding for surface codes, this work opens new directions of research to design new decoders, handle different coding schemes, managing on-chip vs. off-chip trade-offs, and so forth.




\section{Background}
\label{bkg}
\subsection{QEC Overview}
We refer the reader to prior resources on general quantum computing background~\cite{mike_ike_2020,ding2020quantum}
, and limit ourselves to Quantum Error Correction (QEC) overview here.
QEC uses redundancy along with fast and accurate classical processing capability to improve the observed error rates of qubits, thereby improving the fidelity of quantum applications which execute on them.
QEC encodes each logical qubit into a block of physical data qubits.
Further, ancilla qubits are appropriately entangled with each block of data qubits to obtain information about the errors on the data qubits.
The ancilla qubits are repeatedly measured to produce classical error signature bits (called `syndromes'), without destroying the quantum state of the corresponding data qubits.
In fact, the act of ancilla measurement discretizes the data qubit errors into a set of Pauli errors --- i.e., the errors on the data qubits will be an X (bit-flip), Z (phase-flip), Y (both bit and phase flips), or I (no change).
A decoder is then used to try to decipher the error signature and identify the location and types of errors occurring on the data qubits.
This information is then used to add the appropriate corrections to the data qubits. 
If the decoding is performed accurately, then the quantum execution will continue error-free, potentially allowing large-scale quantum problems to be effectively tackled.
If the physical qubit error rates are lower than some threshold (which depends on the error code and the decoder), then increasing the size of the physical data qubit block, which maps to a logical qubit, will monotonically decrease the logical error rates~\cite{9906129}.

\begin{figure}[t]
\centering
\includegraphics[width=0.7\columnwidth,trim={0.25cm 0.25cm 0.025cm 0.05cm},clip]{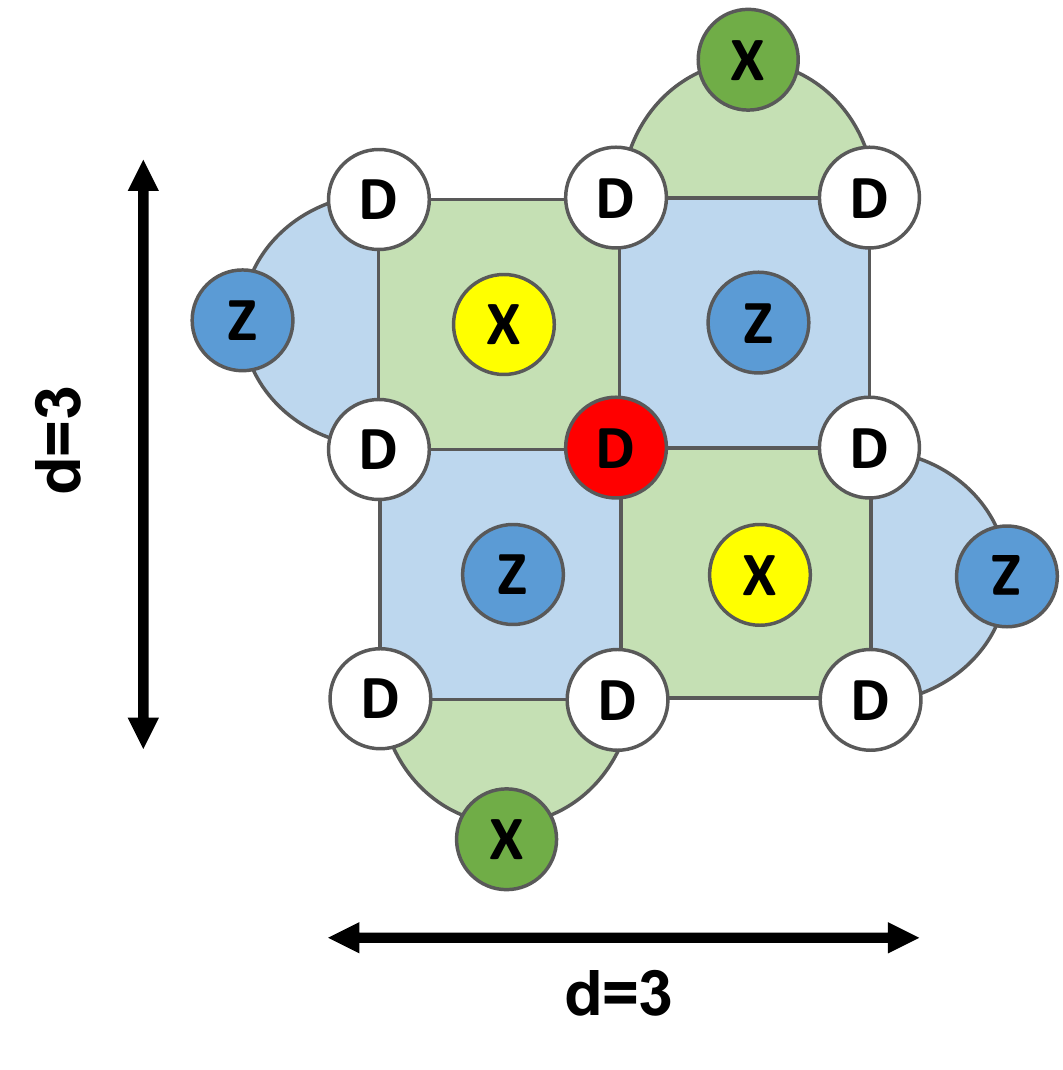}
\caption{
 Distance 3 rotated surface code detecting a Z error on the central data qubit D (highlighted in red) by flipping the neighboring X ancillas (highlighted in yellow).
}
\label{fig:clique_bkg}
\end{figure}

\subsection{Surface Codes}
\label{b:surface}
Surface codes are among the most promising QEC codes, at least for the immediate future, since they have high thresholds allowing for nearly 1\% physical qubit error~\cite{Fowler_2012}.
Further, they require only nearest neighbor physical connectivity --- they encode each logical qubit into a 2-dimensional lattice of alternating physical data and parity (ancilla) qubits, and are therefore amenable to the practical quantum topologies of today~\cite{Lilliput}.
The number of physical data and parity qubits per logical qubit grows quadratically with the code distance `d'.
The code distance is instrumental in setting logical error rates, since the shortest physical error chain that fundamentally cannot be corrected in surface codes is a sequence of `d' errors.
Note that chains shorter than length `d' could also lead to errors if the decoding is performed incorrectly which motivates the use of highly accurate heavy-weight decoders.

An example of surface code for a code distance of `d=3' is shown in Fig.\ref{fig:clique_bkg}.
The white circles `D' represent the data qubits, while the green `X's and blue `Z's are parity/ancilla qubits.
An error which discretizes to a `Z' error on the highlighted red qubit D would be detected by the diagonally adjacent `X' parity qubits, which are highlighted in yellow.
Similarly, `X' errors will be detected on the `Z' parity qubits, and `Y' errors will be detected on both.
These errors are detected through stabilizer circuits, which entangle the parity qubits with the data qubits - more details can be found in previous work~\cite{Fowler_2012}. 

Some points to note:
First, the surface code shown in the figure is known as the rotated surface code, which is a more compact representation and reduces the total physical qubit and gate overheads and is thus preferred.
Second, the corner data qubits and some edge parity qubits have fewer connections compared to other data / parity qubits respectively, which leads to some minimal differences in how they are processed.  
Third, the parity qubits can produce incorrect error signatures if other errors such as temporally transient measurement flips occur. 
Additional redundancy over time (by combining multiple rounds of measurement) is required for robustness against these  measurement error scenarios.

\subsection{Scalability Constraints for Off-chip QEC decoding}
Decoding and error correction should be performed in real-time primarily for two reasons.
First, even though errors can be commuted through many gates (refer Fig.~12 in ~\cite{Fowler_2012} for examples), meaning that those corrections can be applied after the fact, errors cannot commute through T-gates without some conditional S-gate corrections. 
And these corrections are dependent on the history of error signatures before the T gate (Fig.~30 in ~\cite{Fowler_2012}). 
Second, even in the absence of T gates, the logging of syndrome bits and their frequent updates is an arduous task with high memory overheads for larger circuits.
Thus, error decoding is expected to be performed in real time and corrections are to be applied every cycle or at least on every cycle prior to T gates.

It is worth pointing out that quantum computations can technically be stalled prior to T gate execution but unless decoding bandwidth is provisioned to be sufficiently greater than the average rate of syndrome generation, the decoding backlog~\cite{NISQ+} will continue to accumulate every cycle. We dive deeper into this in Section \ref{BWStall}.

Thus, on the one hand, online decoding has tight latency constraints.
On the other hand, the amount of error signature data which needs to be processed per-qubit, every cycle, can grow cubically with the code distance ($d^2$ with qubits, and an additional $d$ with measurement rounds).
Thus, there is clearly a scalability challenge.
Not only does the error signature data grow considerably for more qubits and lower logical error rates, but the decoder complexity to accurately decipher these signatures also has high resource cost.

Off-chip decoding, in which the decoding is performed at room temperature will suffer a bandwidth bottleneck due to the multiple Gbps of syndrome transmission bandwidth required ~\emph{per logical qubit}~\cite{AFS}.
Provisioning for such considerable bandwidth is a serious scalability challenge due to limited I/O wiring for superconducting quantum devices.
\rev{Today, we utilize per-qubit coaxial cables to communicate from the room-temperature QC interface to qubits located within the dilution refrigerator~\cite{XQSim}.
Unfortunately, as the number of qubits increases, and as we begin to support online QEC decoding as described above, this approach significantly suffers from space limitations and huge thermal loads on the cryogenic stage~\cite{Krinner_2019}.
The cross-temperature data transfer through the co-axial cables, dissipates larger heat to the cryogenic side, which is very sensitive to thermal variations and has limited cooling capacity~\cite{XQSim}.}
Over-clustering of I/O wires can also lead to  leakage issues which can further worsen error rates.
\rev{Alternatives to off-chip decoding are discussed in Section \ref{cryo-decoding}.}

\subsection{Cryogenic Classical Hardware}

Quantum devices within dilution refrigerators  require cryogenic classical control to perform numerous functions. 
The controllers handle signal generation to run pulses on each qubit as well as qubit measurement readout and their propagation out of the fridge.
One technology suited to the cryogenic environment is Single Flux Quantum (SFQ), which (despite its name) is classical logic implemented in superconducting hardware.
\rev{SFQ is a magnetic pulse-based fabric with switching delay of only around 1ps and switching energy consumption of $10^{-19}$J.
The switching action consumes 2-3 orders of magnitude lower energy than cryo-CMOS devices.
They are made of Josephson Junctions, which are superconducting devices that exhibit the Josephson effect---indefinitely long current without any applied voltage.
Further, superconducting microstrip transmission in SFQ allow it to transmit at half the speed of light and without dispersion or attenuation. 
The combination of these properties allows for for high speed processing of digital information.}
SFQ logic has been touted as a good candidate for achieving energy efficient and high performance circuits~\cite{8351603}.
More recent versions of SFQ logic family include (Energy-efficient Rapid) ERSFQ~\cite{ERSFQ}. 
Synthesizing for SFQ has different challenges and constraints compared to CMOS and therefore requires its own family of EDA tools~\cite{8351603}.
\rev{These tools are designed to reduce the complexity of the final synthesized and mapped circuits, in terms of total area and Josephson junction count.
This is achieved by reducing the required path balancing logic count for realizing these circuits (these circuits are required to be fully path balanced).
More details in prior work~\cite{NISQ+,8351603,ERSFQ}.}

\subsection{\rev{Cryogenic support for decoding}}
\label{cryo-decoding}

\subsubsection{\rev{NISQ+}}
\rev{The NISQ+~\cite{NISQ+} is entirely an on-chip decoder, i.e., all decodes are handled in the cryogenic domain. Hence it has no off-chip decoding bandwidth and thus entirely alleviates the bandwidth bottleneck. However, it tries to correct even the worst-case errors and thus its design is fairly complex, meaning that it suffers from a resource bottleneck in terms of area, power and thermal  constraints. 
Given the tight resource constraints in the cryogenic domain, it is challenging to scale to large code distances and/or a large number of logical qubits. 
Further, to alleviate some of the resource bottleneck, the decoder is designed in approximate fashion and does not support the correction of measurement errors.
The above discussion on cryogenic decoding suffering a resource bottleneck is broadly applicable to recent/concurrent proposals as well, such as QECOOL~\cite{QECOOL}, QULATIS~\cite{qulatis} and XQSim~\cite{XQSim}.
We show detailed cost comparisons to NISQ+ in Section \ref{e:hw_cost}.}

\subsubsection{\rev{AFS}}
\rev{Prior work, AFS~\cite{AFS} primarily proposes an off-chip decoder that reduces hardware cost through resource sharing.
Additionally, AFS incorporates syndrome compression to reduce the off-chip decoding bandwidth.
We show quantitative comparisons to Clique in Section \ref{e:AFS} but provide brief discussion here.
While compression is attractive when there are no errors (i.e., when the error signatures are all zeros), it is less effective when non-zero elements are present in the error signature, since the bits required to encode these non-zero bits grow quickly. 
We show that Clique is able to reduce bandwidth more than 1000x compared to AFS.
This is because the Clique decoder is able to handle all error signatures that have disconnected non-zero elements and only has to go off-chip otherwise. 
Further, the AFS compressor has to be implemented on-chip in the cryogenic domain. 
The hardware implementation of this compressor (not discussed in the paper) can be rather complex and can itself have a high resource overhead. 
Its power/thermal/area consumption is likely to be a limiter for scalability, constraining the number of logical qubits in the system. 
Finally, our proposals for statistical off-chip bandwidth allocation and decoding overflow based execution stalling are also applicable to AFS since its off-chip bandwidth is dependent on the compression effectiveness in each cycle.}

\section{Motivation: BTWC for QEC Decoding}
\label{motive}
Better Than Worst-Case design is a design philosophy that decouples design for correctness and design for performance and is often pursued in both research and industry, especially in the context of handling classical errors such as those stemming from process variation, timing guard bands and thermal fluctuations~\cite{austin2005opportunities}.
Traditional worst-case design constructs systems with guarantees of correctness and robust operation, which often results in other limitations, such as performance or high resource overhead. 
Better Than Worst-Case designs, on the other hand, achieve high performance and/or low resource cost in the common / average-case and can be designed to gracefully fall back onto the robust worst-case design for the uncommon / worst-case. 

We find this philosophy directly applicable to quantum error correction, simply because the quantum error codes and decoders are provisioned to be robust against very rarely occurring errors so that the target logical error rate is achieved (which is often many orders of magnitude lower than the physical error rate).
In these designs, failure is expected to occur only in scenarios which are beyond the capability of the code or the decoder, such as when a chain of errors occurs with length longer than code distance `d' (as discussed in Section \ref{bkg}).
It is intuitive that if the physical error rate per qubit is `p', then the probability of a chain of `k' errors occurring is proportional to $p^k$ which is clearly $p^{(1-k)}$ times less likely than a single error.

\begin{figure}[t]
\centering
\includegraphics[width=0.98\columnwidth,trim={0cm 0cm 0cm 0cm},clip]{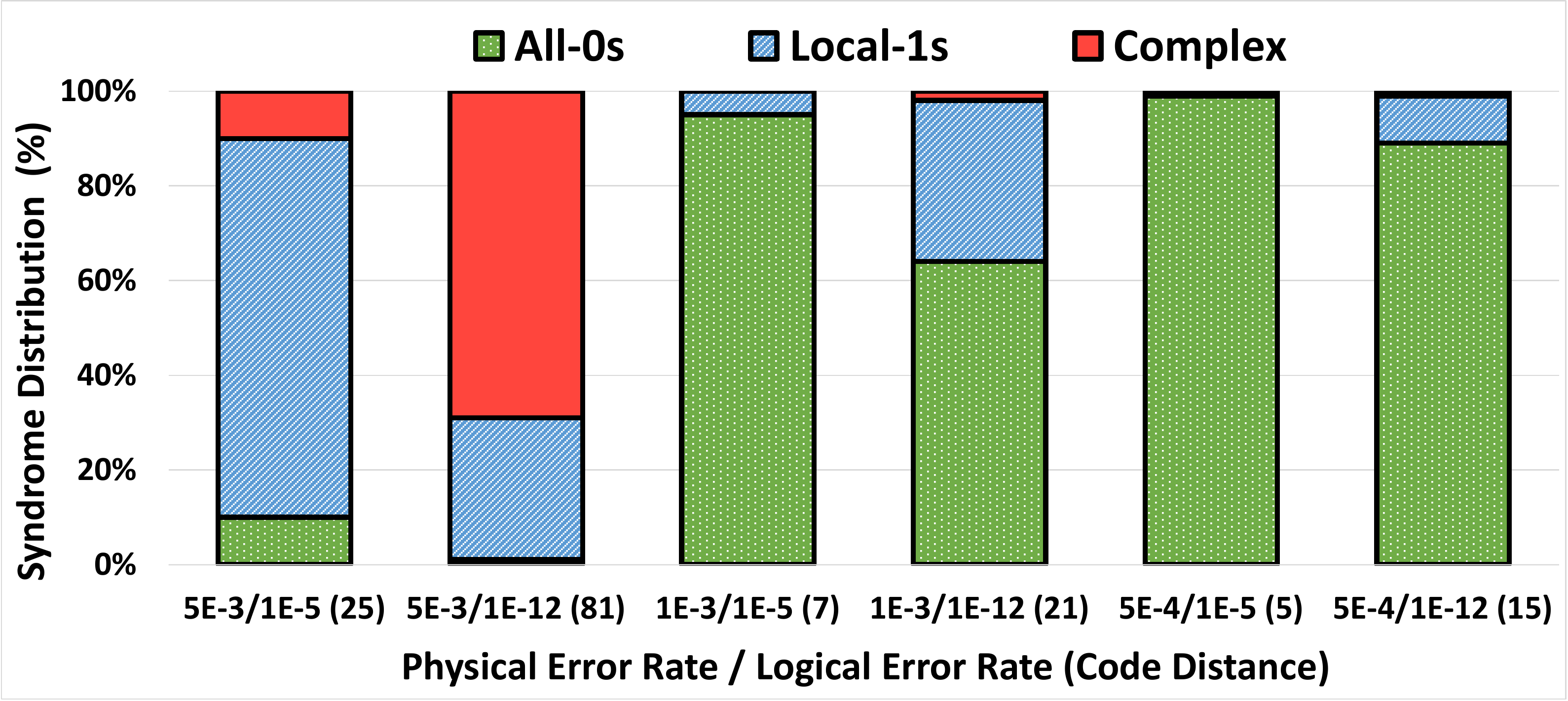}
\caption{
 QEC error signature distributions for various physical and logical error rates (for one logical qubit). When physical error rates are low and/or when code distance is low, the All-0s error signature fraction is generally high. The trivial Local-1s error signatures are fairly significant except in scenarios when physical error rates are low but target logical error rates are very high. The Complex error signatures are fairly rare except when physical error rates are very high and target logical error rates very low.
}
\label{fig:clique_motive}
\end{figure}

This is  reflected in Fig.\ref{fig:clique_motive} which shows QEC error signature distributions obtained over a billion trials for various target physical and logical error rates for one logical qubit (the noise model and simulation methodology are discussed later in Section \ref{sec:method}).
The columns in the figure show a particular combination of a physical qubit error rate, a target logical error rate,  and the code distance required to achieve this. 
We show two logical error rates: 1E-5 and 1E-12. 
The former is suited to near-term advantage in applications like variational algorithms for molecular chemistry, while the latter is suited to long-term targets like search and factorization.
The physical error rates chosen are: 5E-3, 1E-3 and 5E-4.
The first is just below the error threshold for surface codes while the others are improvements we would expect to see over the next decade.
Clearly the code distance needed at the higher physical error rates are high and vice-versa.
Each column shows a breakdown of the error signature distributions, which are described below:

\circled{a}\ The green portions are the error signatures which are all zeroes, i.e. no error was detected in those cycles.
When physical error rates are low and/or when code distance is low, the All-0s fraction is higher since errors are less likely to occur and/or there are a lesser number of physical qubits in the qubit block and thus lower errors in the block.

\circled{b}\ The blue portions (Local-1s) are the error signatures in which errors occur, but there are no error chains with length $\geq 2$ --- all errors are isolated, making decoding fairly trivial (more on this in Section \ref{clique}).
The blue portions are fairly significant, except in Columns 3 and 5, which are scenarios in which physical error rates are low but target logical error rates are rather high (which is not very likely to be the case as we try to push the quantum frontiers).
The relative ratios between the different instances can be inferred similar to the All-0s scenario.

\circled{c}\ The red portions (Complex) are the syndromes in which chains of errors occur --- these are more complicated to decode and can require the full capabilities / resources of complex decoders. 
Note that the red fraction is fairly low (almost negligible in 4 out of 6 scenarios) except in Column 2, which showcases a rather impractical physical-logical error ratio conversion and code distance.

The takeaway from the above is that, in most practical scenarios, a high fraction of the syndromes ($> 90\%$) have error signatures that are trivial to decode and do not exercise the (full) capabilities of the decoder (i.e., the green and blue portions). This clearly motivates a BTWC system design to handle these trivial QEC decodes.

\rev{
From a practical standpoint, in the context of superconducting transmon qubits (which are most suited to surface codes), ~\cite{IBM_Blog1}  discusses physical 2-qubit gate error rates touching 1E-3 on IBMQ Prague and a future envisioning 2E-4.
Further, ~\cite{IBM_Blog2}  discusses scaling to 10K qubits over the next decade.
A logical error rate of 1E-12 is achievable from a physical error rate of 5E-4 with a code distance of 15, which corresponds to roughly 500 physical qubits per logical qubits. 
Thus, running 10+ logical qubits at a logical error rate nearing 1E-12 is a target which could be within reach in the next decade.
Other technologies such as atoms and ions have shown consistent physical error rates of 1E-3, so they are also equally promising.}

\section{Clique Decoder}
\label{clique}

An overview of the Clique Decoder, as part of the overall BTWC QEC architecture, was shown in Fig.\ref{fig:clique_overview}. 
In this section, we dive into its functionality and design, and illustrate some decoding scenarios.

\subsection{Functionality}

The Clique decoder achieves the following functionality:

\circled{a}\ It identifies All-0s signatures and handles them on-chip. No corrections will be applied in these scenarios.

\circled{b}\ It identifies trivial Local-1s signatures and handles them on-chip. Correcting these only involves the manipulation of the data qubits which are nearest neighbors to the erring parity qubits.

\circled{c}\ It identifies Complex signatures and raises a flag to allow them to flow off-chip to a traditional high-cost robust decoder.

\circled{d}\ It neutralizes a high percentage of measurement errors by combining syndrome data from adjacent rounds of measurement (any number of rounds can be incorporated into the design, though more rounds implies greater cost).

\begin{figure}[t]
\centering
\includegraphics[width=0.95\columnwidth,trim={0cm 0cm 0cm 0cm},clip]{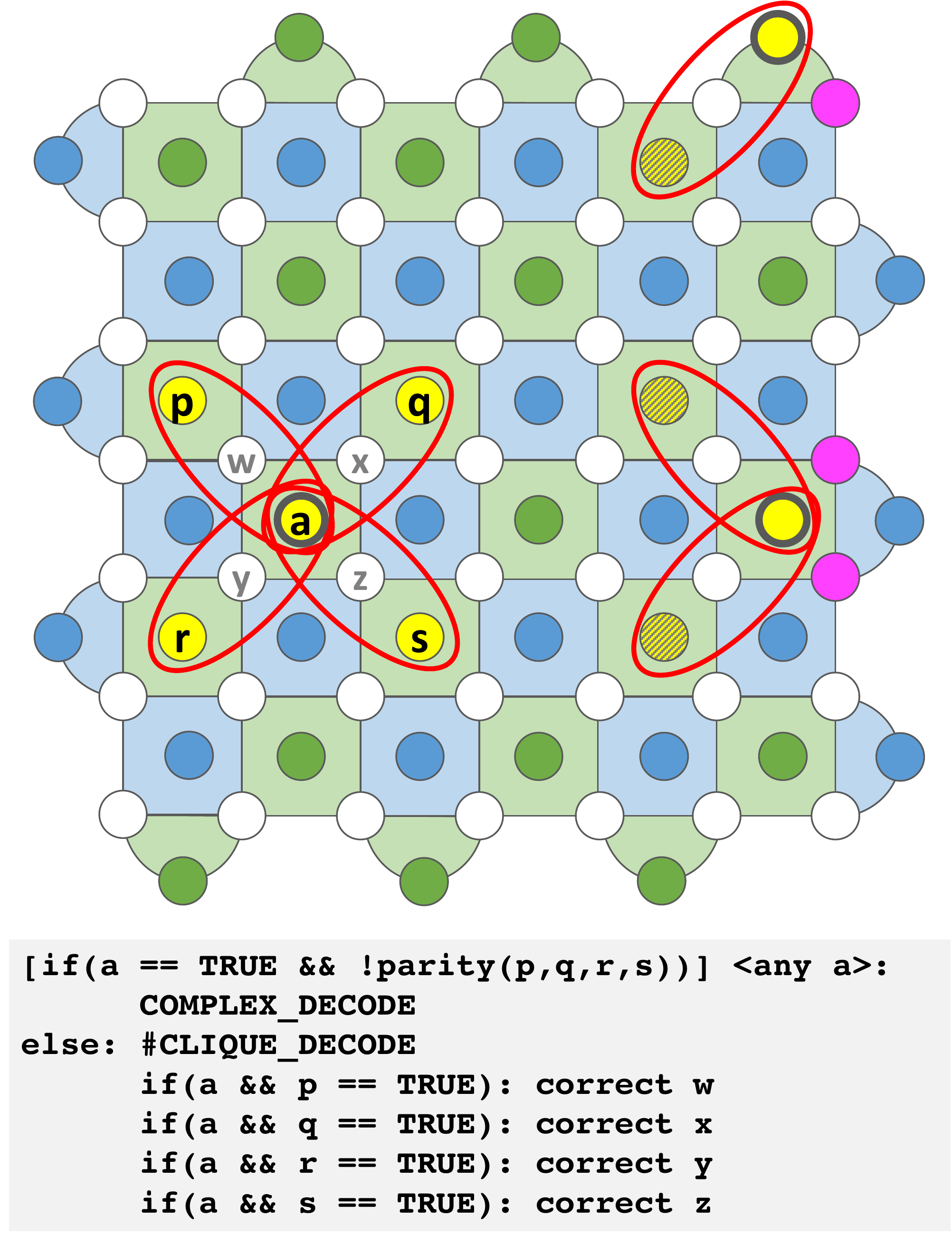}
\caption{
The Clique decoder is designed to analyze the syndromes in every local clique and decide if the error signature is trivial and can be locally handled or is complex and needs to be passed to the off-chip complex decoder. Triviality is decided by calculating the parity of erring ancilla in the surrounding neighborhood of another erring ancilla. If trivial, the appropriate correction is applied to the data qubit which neighbors the erring ancilla. Pseudo-code is shown to the bottom of the figure.
}
\label{fig:clique_design}
\end{figure}

\subsection{Design for Detections, Decisions and Decodes}
To achieve this functionality, the Clique decoder simply analyzes the syndromes from every local clique of ancilla qubits, as shown in Fig.\ref{fig:clique_design}.
The figure shows a $d=7$ physical mapping for a single logical qubit. 
Observe the ancilla qubits highlighted in yellow and labeled `a', `p', `q', `r', `s'.
For every such ancilla qubit `a' that detects an error (creating an `active' clique), the clique decoder checks the parity of errors on the surrounding neighbor ancilla qubits of the same type (i.e., `p'-`s').
If the parity of the neighbors is even (i.e., none of them are set or two of them are set) \emph{for any active clique}, then the decoder deems this to be a complex decode that should be sent off-chip.
If the parity of the neighbors set is odd (i.e., one or three of them are set)  \emph{for all active cliques}, then the decoder deems this to be a trivial decode which can be handled on-chip.
For example, in the event that only the ancillas `a' and `p' are set and `q'-`s' are not, then the data qubit `w', which is their neighbor, has to be corrected. 
The logic performed is described in the pseudocode at the bottom of Fig.\ref{fig:clique_design}.

A couple of points to note.
First, the discussion above focuses on blue `Z' type ancillas, but the same applies for green `X' type ancillas as well.
Second, the corner and edge qubits are treated a bit differently depending on their neighborhoods. 
Two examples of this are shown in the figure. The respective `a' parity qubits in these examples have a thicker outline.
For the example at the top right, the clique has only one neighbor (i.e., 1+1). This clique is a special case and is always trivial to decode, even if the neighborhood parity is even.
An even parity here (i.e., neighbor is unset, indicated by hatch)  will simply mean that the data qubit in pink, adjacent to the active syndrome, has to be corrected and complex decode is not required.
For the example at the center right, the clique has two neighbors (i.e., 1+2). 
This clique is also a special case and is trivial to decode if both neighbors are unset (even though it would mean even parity), apart from the usual odd parity case.
If both neighbors are unset (indicated by hatch), then one of the two data qubits in pink, adjacent to the active syndrome, has to be corrected and complex decode is not required.
Interestingly, flipping either of the two data qubits in pink is equivalent and sufficient to correct the error! This is because two corrections are equivalent if they differ by a stabilizer.
We leave interested readers to learn more about such scenarios from ~\cite{Fowler_2012} (Fig.3 in the paper and related text).
Other instances such as these are distributed over the corners and edges. 

\begin{figure}[t]
\centering
\includegraphics[width=0.95\columnwidth,trim={0cm 0cm 0cm 0cm},clip]{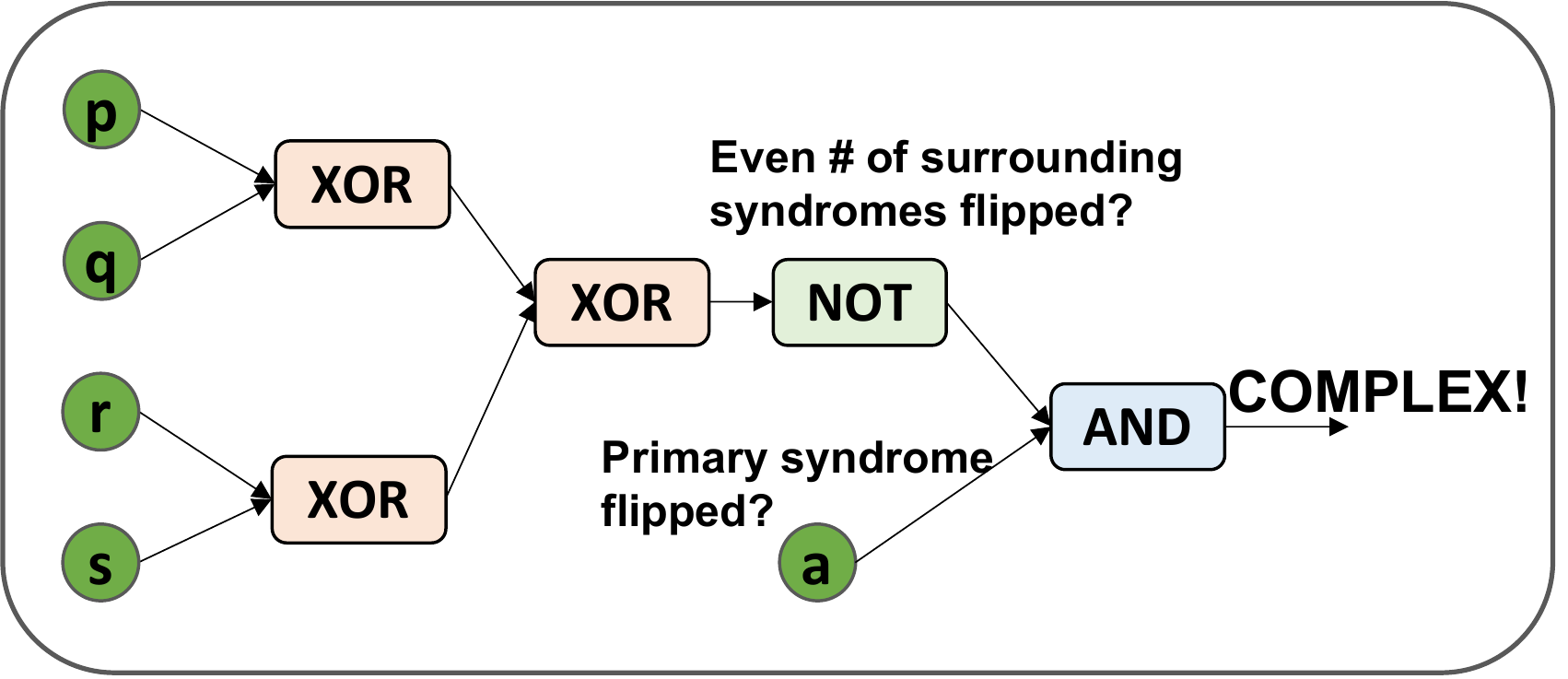}
\caption{
 The Clique decoder's decision logic (to decide trivial vs. complex) is a small number of logic gates per clique and is especially suited to on-chip implementation.
}
\label{fig:clique_detect}
\end{figure}

The Clique decoder is especially favorable for on-chip implementation due to its lightweight hardware - it is simply requires a few combinational logic gates per clique.
The gate-level decision logic to decide whether a syndrome is trivial or complex is shown in Fig.\ref{fig:clique_detect}. 
The XOR gates are used to find the parity of the neighborhood and this is combined with the check on whether the clique is active or not.
This logic is performed on each clique.
Note that if even a single clique raises a COMPLEX output, then the syndromes are passed off-chip.
Furthermore, the error correction control logic is simply an AND gate (of the ancillas), which feeds into a conditional gate in the quantum circuit and is not shown at the circuit-level but is described in the pseudocode in Fig.\ref{fig:clique_design}.

\begin{figure}[t]
\centering
\includegraphics[width=0.95\columnwidth,trim={0cm 0cm 0cm 0cm},clip]{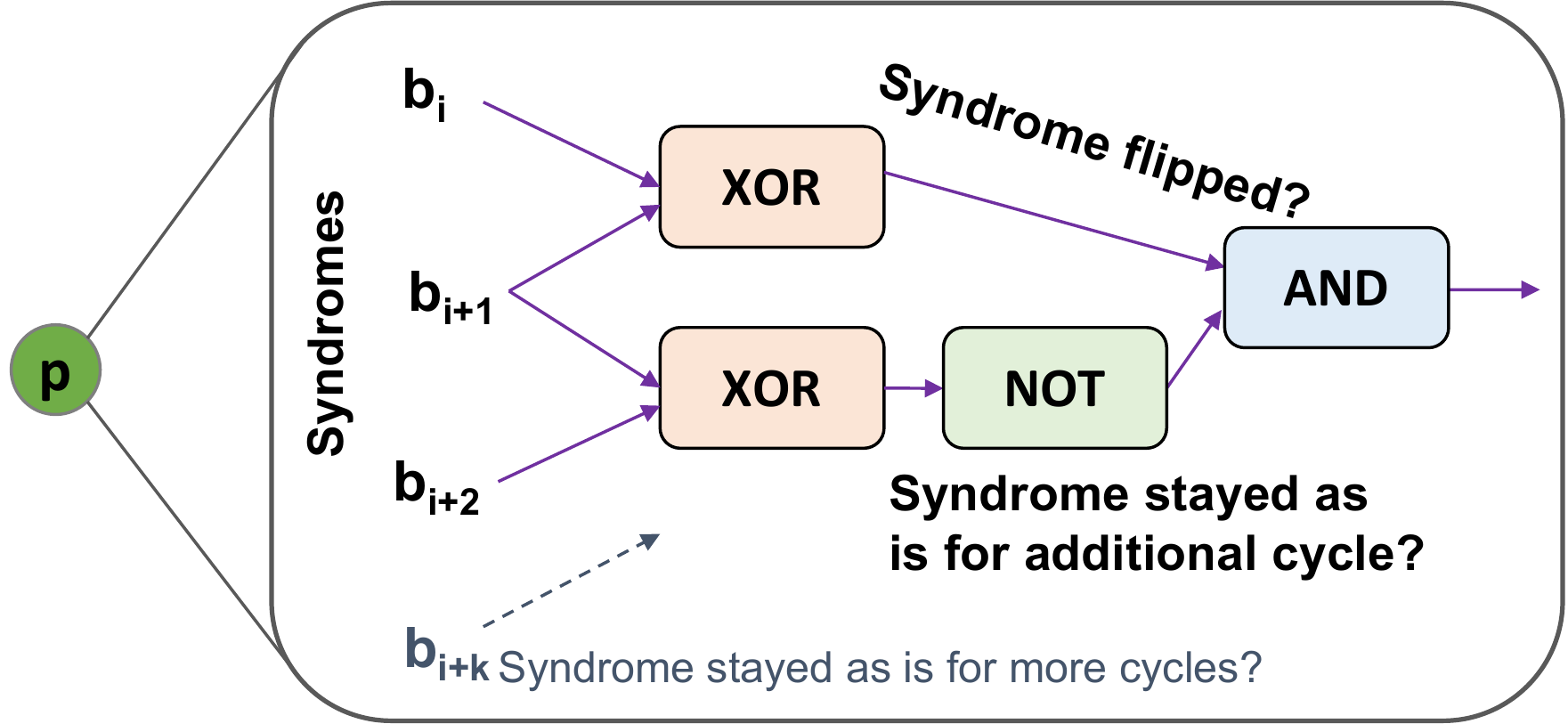}
\caption{
 Errors occurring when ancilla qubits are measured are handled by evaluating multiple rounds of measurements and by only considering errors that stick over the measurement rounds. The Clique decoder uses only two measurement rounds, but more rounds can be added for higher accuracy at additional hardware cost.
}
\label{fig:clique_measure}
\end{figure}

\subsection{Handling Measurement Errors}

\begin{figure*}[t]
\centering
\includegraphics[width=0.95\textwidth,trim={0cm 0cm 0cm 0cm},clip]{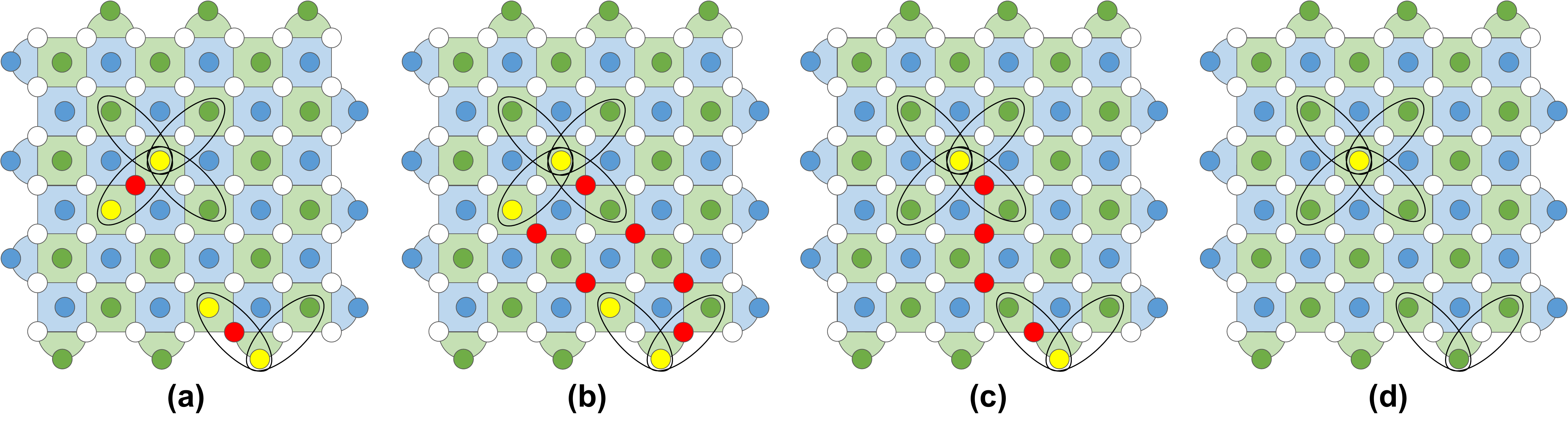}
\caption{
 Different error scenarios and their decoding. (a) A trivial error signature in which two local data qubit errors have occurred (red) and are reflected in the nearby ancillas. The Clique decoder detects this and applies the appropriate correction. This is the same correction that would be applied by a complex decoder (b) It is possible that the same error signature was caused by a more complicated coordinated sequence of data qubit errors. However, this is extremely unlikely - this scenario would cause both Clique as well as complex decoders to fail. (c)  A complex error signature caused by a shorted sequence of coordinated data qubit errors. Clique will detect this complex signature and hand it over to the complex decoder. (d) An error signature caused by a measurement error. If the error persists beyond the measurement rounds, then Clique will detect this to be a complex error signature and again hand it over to the complex decoder.
}
\label{fig:clique_scenarios}
\end{figure*}

Next, we discuss how measurement errors are handled.
In Fig.\ref{fig:clique_detect}, if we were \emph{not} concerned about measurement errors, then `a', `p', `q', `r', `s' in the figure would simply be the syndromes directly measured from the ancilla parity qubits.
But in the presence of measurement errors, particular qubits could randomly flip on some measurement cycles as briefly noted in Section \ref{b:surface} (and illustrated in Fig. 2 in \cite{Fowler_2012}).
To avoid this, parity qubits are measured over multiple measurement rounds and those errors which stick across these rounds are considered data errors, while those which disappear are ignored as measurement errors (or self-corrected data errors).
Clearly, the more rounds of measurement the better, similar to the spatial code distance argument. 
For the Clique decoder, our primary design uses two rounds of measurement but more rounds can be added at additional hardware cost.
This is illustrated in Fig.\ref{fig:clique_measure}.
In the figure, the subscript `i' indicate the cycles over which the measurements are captured and logically combined together.
The XOR is indicative of a syndrome flip over a cycle, and the overall logic checks if the initial flipped state persists across the rounds.
If the flip has persisted over the rounds, then the error is likely to be interpreted as complex --- the intuition for this is discussed in Section \ref{intuition}.

\subsection{Decoding Intuition and Examples}
\label{intuition}

Before we discuss the decoding decisions made by Clique, we first provide some decoding intuition that is fundamental to any decoder.
There are usually multiple error configurations possible (caused by data, measurement, etc.) that could have produced a particular error signature.
Any decoder usually chooses the error configuration with the minimum number of errors that could have produced the particular error signature.
The reason for this follows the discussion in Section \ref{motive}, that the occurrence of `n' errors is roughly $r*(n-m)$ \emph{orders of magnitude} less likely to occur than `m' errors for $n>m$, under the reasonable assumption that all qubits have similar error rates roughly equal to $10^{-r}$.
If ties occur in terms of multiple error configurations with the same number of errors, then ties can be broken with more specific information about per-qubit error rates. 

With the above intuition, we now show different error scenarios in Fig.\ref{fig:clique_scenarios} and justify that the decisions made by Clique for trivial error signatures are equivalent to those made by a heavy-weight robust decoder, and for more complex scenarios, Clique accurately hands over to the robust decoder.

First, Fig.\ref{fig:clique_scenarios}.a and Fig.\ref{fig:clique_scenarios}.b show the same error signature with two sets of paired erring ancillas, which are shown in yellow.
Fig.\ref{fig:clique_scenarios}.a shows a corresponding error configuration with two data errors (shown in red), while Fig.\ref{fig:clique_scenarios}.b shows an error configuration with six data errors.
Any decoder would be expected to select the error configuration in Fig.\ref{fig:clique_scenarios}.a which is 8 orders of magnitude more likely to occur for a physical error rate of $10^{-2}$.
Clique decides that this error signature is trivial since the two active cliques (i.e., those with `a' ancilla set) both have an odd neighbourhood parity. 
Then, within each clique, the decoder directs the fix on the data qubit which is a neighbor to the erring ancilla, as described in Fig.\ref{fig:clique_design}.
Thus, Clique achieves the same decoding and correction as a high-cost complex decoder.

Next, Fig.\ref{fig:clique_scenarios}.c shows a scenario with two stand-alone erring ancillas in yellow.
The most likely error configuration here is a chain of 4 erring data qubits (assuming that the possibility of  measurement errors have been eliminated over the measurement rounds) as shown in red in the figure. 
This is the likely configuration chosen by a complex decoder.
Clique is unable to handle this decode, but correctly detects this and passes it over to the complex decoder.
It is able to do so because it identifies that there exists at least one active clique with an even neighborhood parity, as described in Fig.\ref{fig:clique_design}.
Importantly, note that the Clique decoder is not required to know that there is more than a single active clique --- all its decodes are local to a clique --- this is key to its lightweight design.

Finally, Fig.\ref{fig:clique_scenarios}.d shows a scenario with a single erring ancilla, which is likely caused by a long-lasting measurement error (there is no ancilla pairing that can lead to a chain of data errors).
Clique views this decode exactly as it viewed Fig.\ref{fig:clique_scenarios}.c. 
Clique identifies that there exists at least one active clique that has an even neighborhood parity, and thus passes to the complex decoder.
Note that, with its myopic local view, Clique cannot differentiate between Fig.\ref{fig:clique_scenarios}.c and Fig.\ref{fig:clique_scenarios}.d (i.e., that one was caused by a sequence of data errors and the other was caused by a measurement error).
But importantly in both cases, Clique detects its decoding incapability and allows more robust handling off-chip.

Note that, while we do not showcase in our previous illustrations, intersecting cliques can also be handled by the Clique decoder---all syndromes errors will be covered in one or more cliques. 
In such a scenario, if any of the cliques deems that its errors are complex (i.e., it is an active clique and its neighborhood has even parity) then the clique decoder will pass over to the off-chip decoder. This will work for any error or errors that occur anywhere over the entire block of physical qubits.
For example, it is okay if diagonally adjacent cliques (which will share one of the 4 ‘leaves’ of the clique) cover the same pair of syndrome errors, because they will both just indicate that a particular data qubit needs to be fixed at the end of the decoding cycle. This is just an AND operation of the syndrome bits, as shown at the bottom of Fig.\ref{fig:clique_design}---it does not matter which clique(s) is/are triggering it.

\begin{figure}[t]
\centering
\includegraphics[width=0.98\columnwidth,trim={0cm 0cm 0cm 0cm},clip]{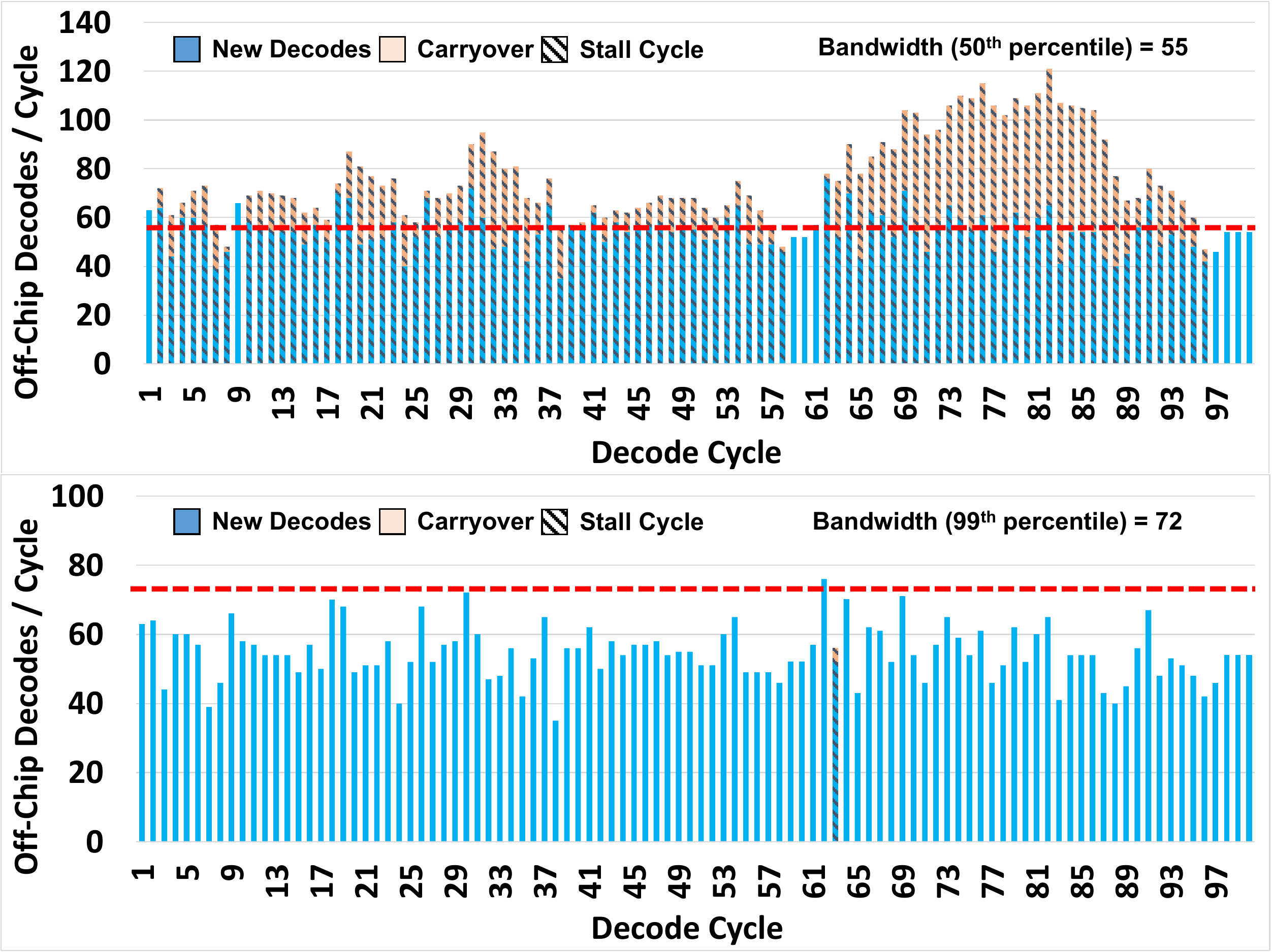}
\caption{
Provisioning off-chip bandwidth for the average off-chip decoding rate is insufficient because it will likely lead to a decoding backlog that cannot be overcome. Provisioning more conservatively allows for optimal BTWC handling with a minimal number of stall cycles. 
}
\label{fig:clique_bw}
\end{figure}

\section{Bandwidth Allocation and Overflow Stalling}
\label{BWStall}

The Clique Decoder can handle common-case trivial decodes, and more complex decodes are handed over to the off-chip decoder.
This requires appropriate provisioning of per-logical-qubit off-chip bandwidth, as well as a fail-safe mechanism in case the provisioning is exceeded. 
This was illustrated in Fig.\ref{fig:clique_overview} and is discussed in detail next.

\subsection{Statistical Off-chip Bandwidth Allocation}
Consider a scenario in which the Clique decoder has a decode coverage of around 95\%.
This means that only about 5\% of the decodes, per logical-qubit, will have to be transferred off-chip for complex decoding.
For a quantum device with 1000 logical qubits, this means that roughly 50 off-chip decodes are generated every cycle.
However, it is insufficient to provision the off-chip bandwidth for this average case scenario which is depicted in the top graph in Fig.\ref{fig:clique_bw}.
The graph shows the 1000 logical qubit system over 100 decode cycles.
If off-chip decode bandwidth is provisioned for around the 50th percentile (= 55 decodes per cycle), then there are many cycles in which the number of newly produced off-chip decodes in that cycle exceed this provisioning --- this is indicated by the height of the blue bar exceeding the red line which represents off-chip provisioning.
Minimally, this means that the next cycle has to be stalled so that the corrections can be applied before subsequent gates are performed on the qubits; a stall cycle is shown via hatched bars in the graph.
However, the stall cycle itself is not free of errors, since qubits are still free to decohere etc. 
Thus, new errors, and potentially new off-chip decodes, are produced even in the stall cycle.
Now the stall cycle has to perform off-chip decoding for both the carryover decodes from the previous cycle (since they will again be avoided by the Clique decoder) as well as the new off-chip decodes, the sum total of which is bound to be even greater than the average provisioning. 
In the graph, the carryover off-chip decodes are shown in orange.
Clearly, the sum of carryovers and new off-chip decodes constantly tends to exceed the off-chip bandwidth provisioning, leading to more than 90 cycles of stalling in a 100 cycle window.
Thus, it is evident that average provisioning leads to a decode backlog problem.

To avoid this, off-chip bandwidth is statistically allocated such that a fairly high fraction of the off-chip decodes can be decoded every cycle.
The bottom graph in Fig.\ref{fig:clique_bw} shows the same system being provisioned for the 99th percentile off-chip bandwidth, which is 30\% greater than the previous scenario.
In this case, the off-chip decodes that are generated in all but one cycle are able to flow through to the complex decoder in the same cycle that they are generated.
Only a single cycle is shown to cause a off-chip decode overflow, leading to a stall in the subsequent cycle.
Further, the bandwidth is comfortably provisioned such that the new off-chip decodes plus the carryovers can be handled in the stall cycle's decoding, thus avoiding an accumulating decode backlog problem.

In evaluation, we explore the stalling vs bandwidth provisioning trade-off in a more fine-grained manner for different physical error rates and code distances.
Finally, it should be noted that lower bandwidth provisioning does not mean under-utilization of the available I/O, it would instead be used to execute more logical qubits in parallel.

\begin{figure}[t]
\centering
\includegraphics[width=0.98\columnwidth,trim={0cm 0cm 0cm 0cm},clip]{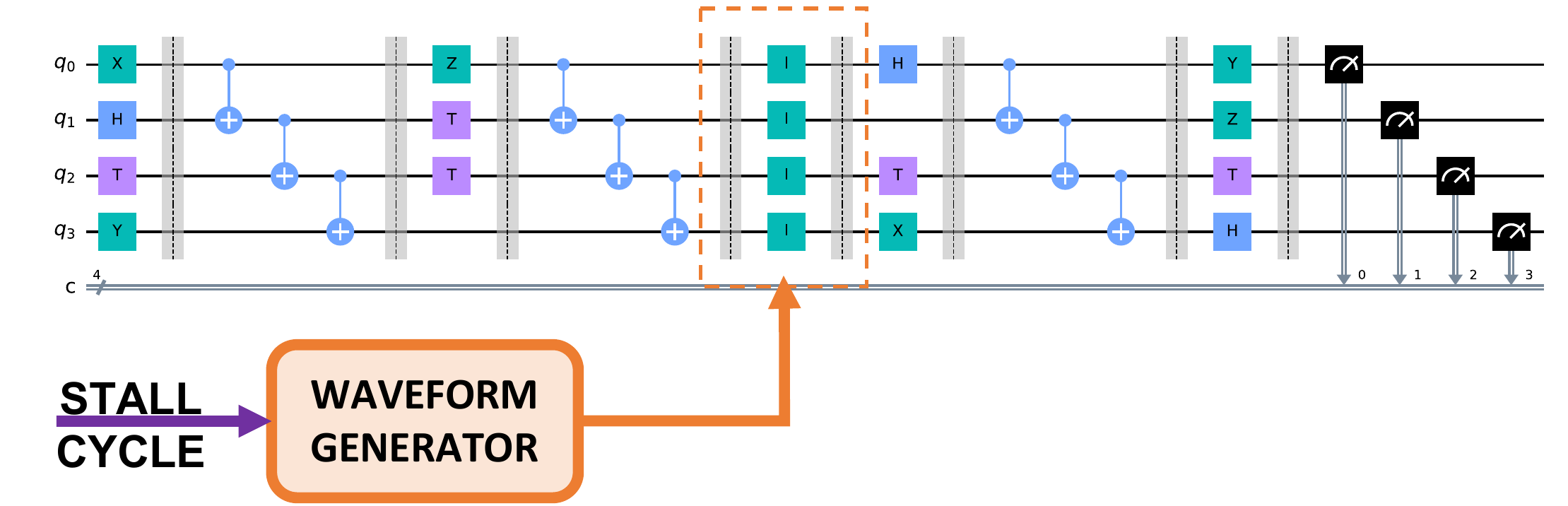}
\caption{
Idle cycle insertion if a stall signal is received.
}
\label{fig:clique_stall}
\end{figure}

\subsection{Decode-Overflow Execution Stalling}

Off-chip decode overflow  is detected in a particular cycle if the number of logical qubits that need off-chip complex decoding exceeds the provisioned off-chip bandwidth. 
In this scenario, a stall cycle has to be generated. 
To do so, a control signal is sent to the Waveform Generator (which sends gate pulses to the qubits every cycle). 
On a stall cycle, no operations are performed on the qubits.
This is illustrated in Fig.\ref{fig:clique_stall} for an example 4-qubit circuit. 
The stall cycle is indicated by the Identity gate being performed over all the qubits and is a fairly trivial operation.

\section{Methodology}\label{sec:method}

\subsection{Simulation Infrastructure and Noise Model}
To benchmark the performance of the Clique decoder, we simulate its action over a billion random cycles of execution. 
This is called lifetime simulation or Monte Carlo benchmarking. 
\rev{Our simulation infrastructure and noise model are similar to that used in prior work~\cite{AFS, NISQ+, VLQ}.}

We construct a simulation environment in which errors are stochastically injected onto the qubits.
The effect of data and measurement errors is then reflected in the per-cycle syndrome measurements from the ancilla qubits.
The error signatures are then sent to the Clique decoder simulator, which decides if it can decode the syndrome or not and, if so, returns the appropriate correction.
If not, corrections are obtained from an implementation of the state-of-the-art Maximum Weight Perfect Matching (MWPM) decoder~\cite{Dennis_2002}.
The correction is then applied, and the surface is checked for logical errors.
The effectiveness of the decoder is evaluated on the basis of its coverage and error rates.
The above simulation is performed for one logical qubit, with varying code distances and physical error rates.

To evaluate the bandwidth allocation and stalling optimizations, simulation similar to the above is performed for 1000 logical qubits, over a million random execution cycles, again for varying code distances and physical error rates.
In this scenario, the fraction of decodes which are avoided by Clique is captured per logical qubit, per cycle.
This is then used to evaluate the trade-off between different statistical provisioning of  off-chip bandwidth and the number of stalls required for the off-chip decoding to keep up with the rate of error generation.

Noise is introduced in the simulation framework described above based on the phenomenological noise model for both data qubit errors and measurement errors~\cite{AFS}.
X-type and Z-type errors are corrected independently, so focusing on either one is sufficient for modeling purposes. 
Our noise model is parameterized by `p ---each cycle introduces errors on the data qubits with a probability of `p' and errors on the syndrome measurement with the same `p' probability.  

\rev{While actual applications running on real devices will not mimic this noise model on a cycle to cycle basis, the average effects are expected to be similar and thus the evaluated decoder benefits will also be similar.
Evaluation can only be performed with probabilistic simulated error models, and not on actual applications on real hardware, because we (meaning the totality of quantum industry and research) do not have large enough devices to run fault tolerant quantum applications on them.}

\begin{table}[t]
\centering
\resizebox{0.85\columnwidth}{!}{%
\begin{tabular}{|l|l|l|l|}
\hline
\textbf{Gate} & \textbf{Delay (ps)} & \textbf{Area ($\mu$$m^2$)} & \textbf{JJ Count } \\ \hline
XOR2  & $6.2$  & $7000$ & $18$ \\ \hline
AND2  & $8.2$  & $7000$ & $16$ \\ \hline
OR2   & $5.4$  & $7000$ & $14$ \\ \hline
NOT   & $12.8$ & $7000$ & $12$ \\ \hline
DFF   & $8.6$  & $5600$ & $10$ \\ \hline
SPLIT & $7.0$  & $3500$ & $4$  \\ \hline

\end{tabular}%
}
\caption{ERSFQ cell library used for decoder synthesis}
\label{tab:SFQ_lib}
\end{table}

\subsection{Cryogenic Hardware Implementation}
The lightweight resource requirements of the Clique decoder make it well suited to cryogenic implementation.
\rev{The Clique decoder is written in verilog and synthesized for SFQ hardware.
The synthesis is performed via a framework incorporating methodologies from SFQMap~\cite{8351603} which is a technology mapping tool for SFQ circuits.}
While there are no fundamental constraints that limit Clique implementation to a specific cryogenic technology, we implement with ERSFQ technology~\cite{ERSFQ} in this work.
The ERSFQ cell library used is shown in Table \ref{tab:SFQ_lib}.
\rev{Logic in SFQ is built from Josephson Junctions (JJ), which are superconducting devices that exhibit the Josephson effect. This  JJ count is shown in the Table for different logic gates.}
More details on how SFQ logic synthesis is performed can be found in previous work~\cite{NISQ+,8351603}.

\section{Evaluation}

\subsection{Clique On-Chip Coverage} 

\begin{figure}[t]
\centering
\includegraphics[width=0.95\columnwidth,trim={0cm 0cm 0cm 0cm},clip]{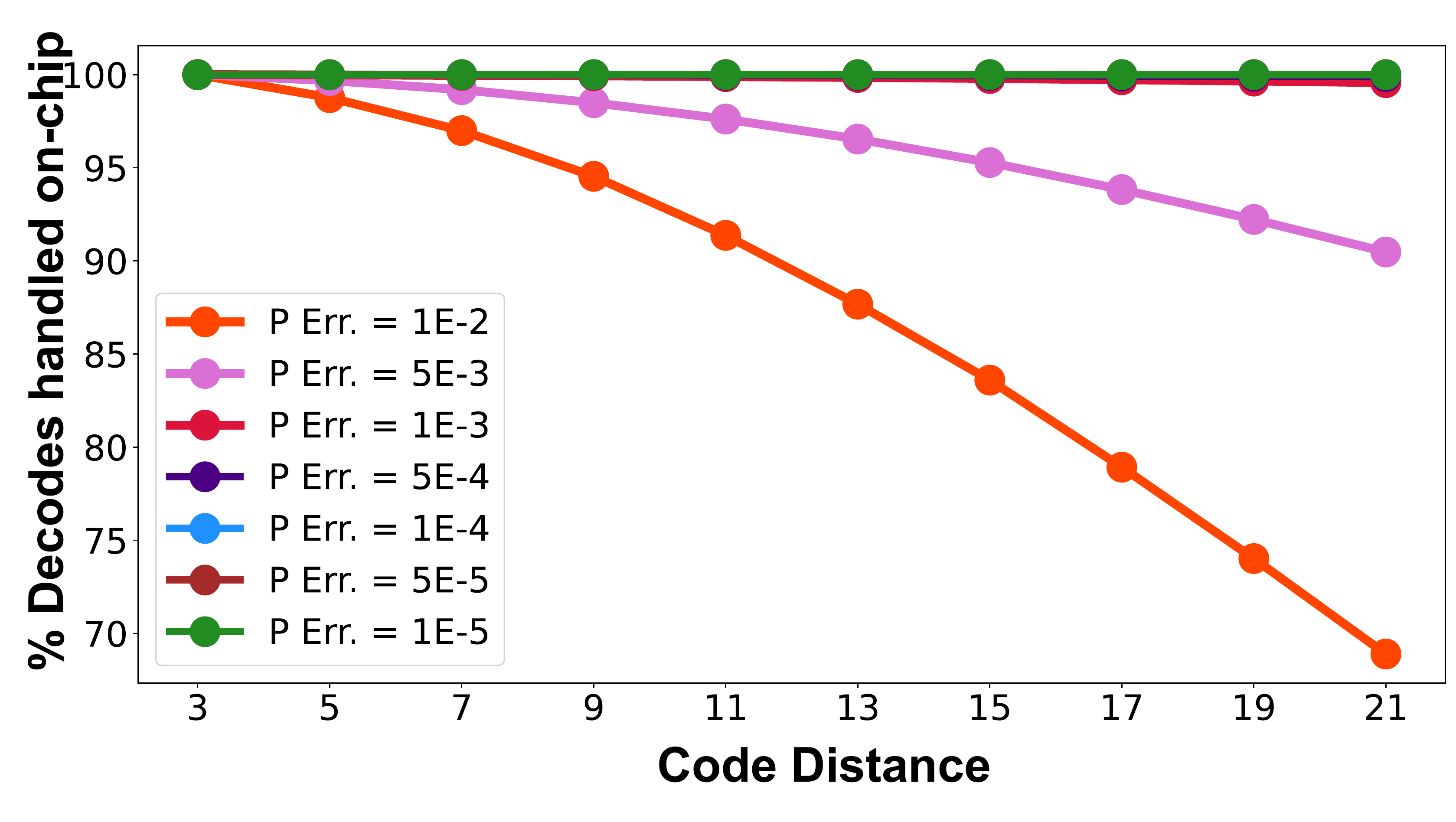}
\caption{
Fraction of decodes that can be handled by Clique without having to go off-chip.
}
\label{fig:clique_coverage}
\end{figure}

First, we evaluate the coverage of the Clique decoder, in terms of the fraction of decodes that can be handled by Clique without having to go off-chip. 
This is evaluated for different physical error rates and code distances and is shown in Fig.\ref{fig:clique_coverage}.
The first takeaway is that the Clique coverage is high (~70\%) even at high physical error rates and high code distances.
This is the scenario that is relatively more challenging for Clique.
For example, at a code distance of 21, there are roughly 1000 sources of error (X / Z data errors and measurement errors) per logical qubit.
At a physical error rate of 1\%, this would mean that an average of 10 errors would occur every cycle.
But many of these errors are still trivial since a coordinated chain of errors is still very unlikely to occur, thus the coverage is still nearly 70\%.
As physical error rate decreases and/or as code distance decreases then the Clique coverage increases further to nearly (but still under)  100\% .
This is intuitive because errors, in general, and complicated coordinated errors, in particular, become increasingly more rare in these scenarios. 
But note that complicated errors \emph{do} exist and \emph{must} be corrected to achieve the target logical error rates.

\begin{figure}[t]
\centering
\includegraphics[width=0.95\columnwidth,trim={0cm 0cm 0cm 0cm},clip]{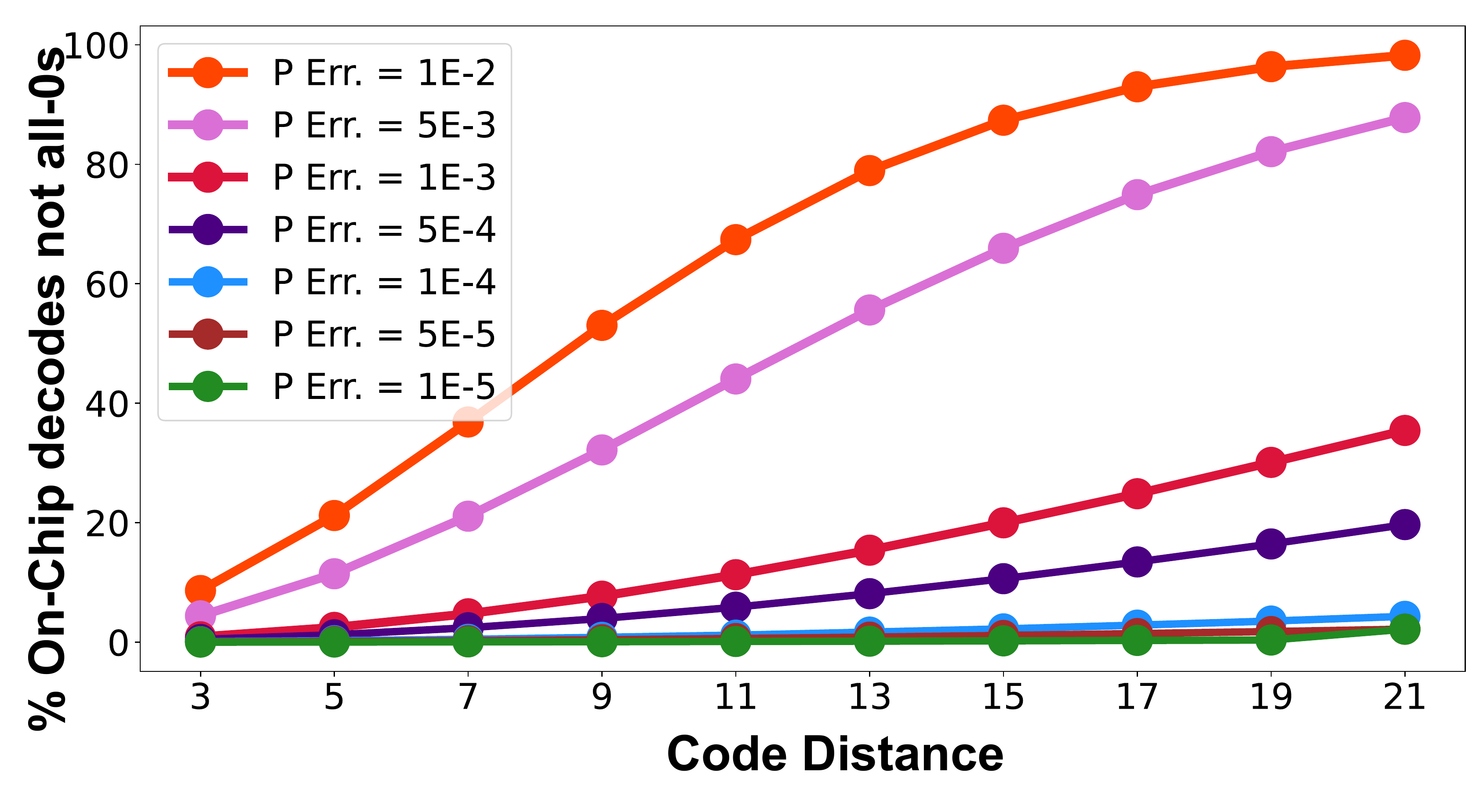}
\caption{
Fraction of decodes which are actually errors but trivially handled by Clique.
}
\label{fig:clique_coverage_non0}
\end{figure}

Next, we evaluate the fraction of decodes that are actually errors, but are trivially handled by Clique.
This is important because, if the error signatures were (nearly) all zeros, then on-chip handling could potentially be simpler than Clique. 
Fig.\ref{fig:clique_coverage_non0} shows this evaluation for the same physical error rates and code distances as before. 
Clearly the fraction of non All-0s decodes that are handled on-chip is nearly 100\% near the surface code error threshold (i.e., highest physical error rates), especially at high code distances.
Thus, going off-chip for all errors that are not all-0s is not sufficient for significant decoding bandwidth reduction, and a better decoder like Clique is required. 
Note that most of the decodes can be All-0s for very low physical error rates, but it is debatable whether surface codes would actually be the code of choice at physical error rates lower than, say, $0.01\%$---other schemes such as Steane codes and Bacon-Shor codes could be better options~\cite{Fowler_2012}. 
But in the foreseeable future of error rates, the benefits of Clique are abundantly clear.
\rev{For any QEC code of choice, there will always be separation between common trivial error signatures and rare complex ones, because this stems from the innate redundancy in encoding schemes.
In fact, the common cases will only become more common as technology improves and error rates decrease, since the likelihood of complex error chains become even smaller.}

\subsection{Comparison against AFS Syndrome Compression}
\label{e:AFS}
\begin{figure}[t]
\centering
\includegraphics[width=0.95\columnwidth,trim={0cm 0cm 0cm 0cm},clip]{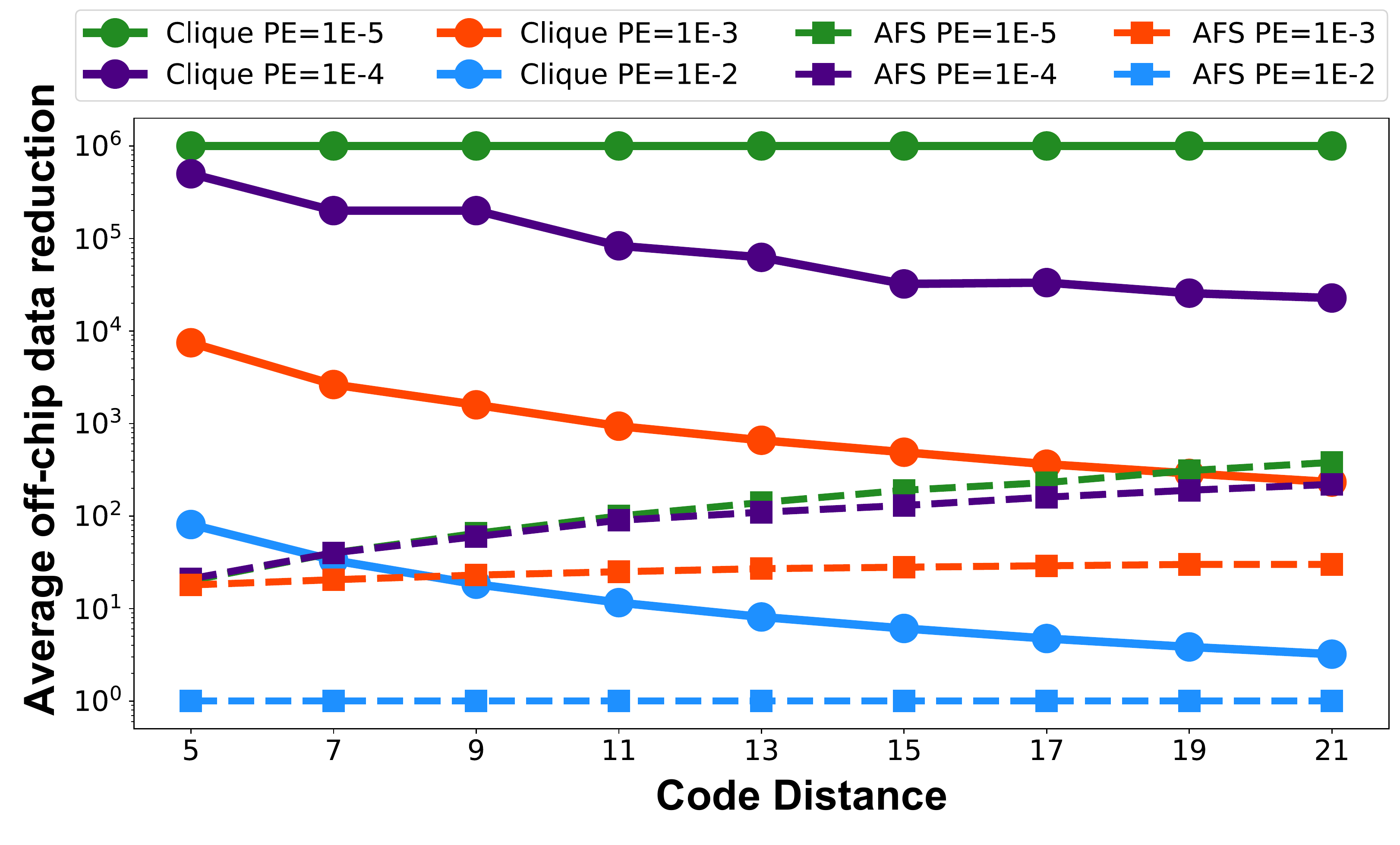}
\caption{
Comparison against Syndrome Compression.
}
\label{fig:clique_afs}
\end{figure}

The decoding bandwidth from error signatures with sparse errors (i.e., less number of 1s) can also be reduced through syndrome compression which has been pursued in prior work~\cite{AFS}.
However, we argue that the Clique decoder is a more attractive solution for multiple reasons.

First, syndrome compression is most attractive when error signatures are entirely (or almost all) zeroes. 
This is because compression techniques usually have high cost to represent the bits that are not zeroes.
For example, AFS adopts a Sparse Representation technique (their most effective technique) that uses a Sparse Representation Bit to indicate if all the
syndrome bits are 0s, which achieves $O(N)$ reduction for an All-0 N-bit syndrome.
However, it uses the indices for all non-zero bits, which grows as $1+O(k*log(N))$ where `k' is the number of non-zero bits.
At low code distances, benefits can be limited because N is low, whereas at higher error rates, benefits can again be limited because `k' is high.
On the other hand, Clique obtains 100\% bandwidth elimination on all instances of error signatures with All-0s or Local-1s.
Quantitative comparison of off-chip data reduction between AFS and Clique (on a log scale) is shown in Fig.\ref{fig:clique_afs}.
Observe that Clique provides many orders of magnitude improvement over AFS (10x-10,000x) for the reasons discussed.
Note that for a given error rate, AFS benefits will increase with code distance but Clique benefits decrease but both eventually saturate.
The Clique saturation benefit is at least an order of magnitude higher than that of AFS. 
AFS's benefits grow initially due to its inherent limitation at lower code distances as discussed prior. 

Second, syndrome compression is not a trivial task and still has to be performed on chip.
To our knowledge, AFS~\cite{AFS} does not provide a hardware implementation of the proposed compression technique. 
These techniques, while effective in theory (and in software), can have high hardware cost.
Sparse Representation hardware cost grows quickly with the number of bits in the syndrome.
AFS, in fact, proposes to implement three techniques and choose between them dynamically based on compression ratios. 
This can be substantially complicated for on-chip implementation.
On the other hand, Clique hardware costs are substantially low and are discussed in Section \ref{e:hw_cost} in comparison to other work with on-chip implementation.

\subsection{Logical Error Rates}

\begin{figure}[t]
\centering
\includegraphics[width=0.99\columnwidth,trim={0cm 0cm 0cm 0cm},clip]{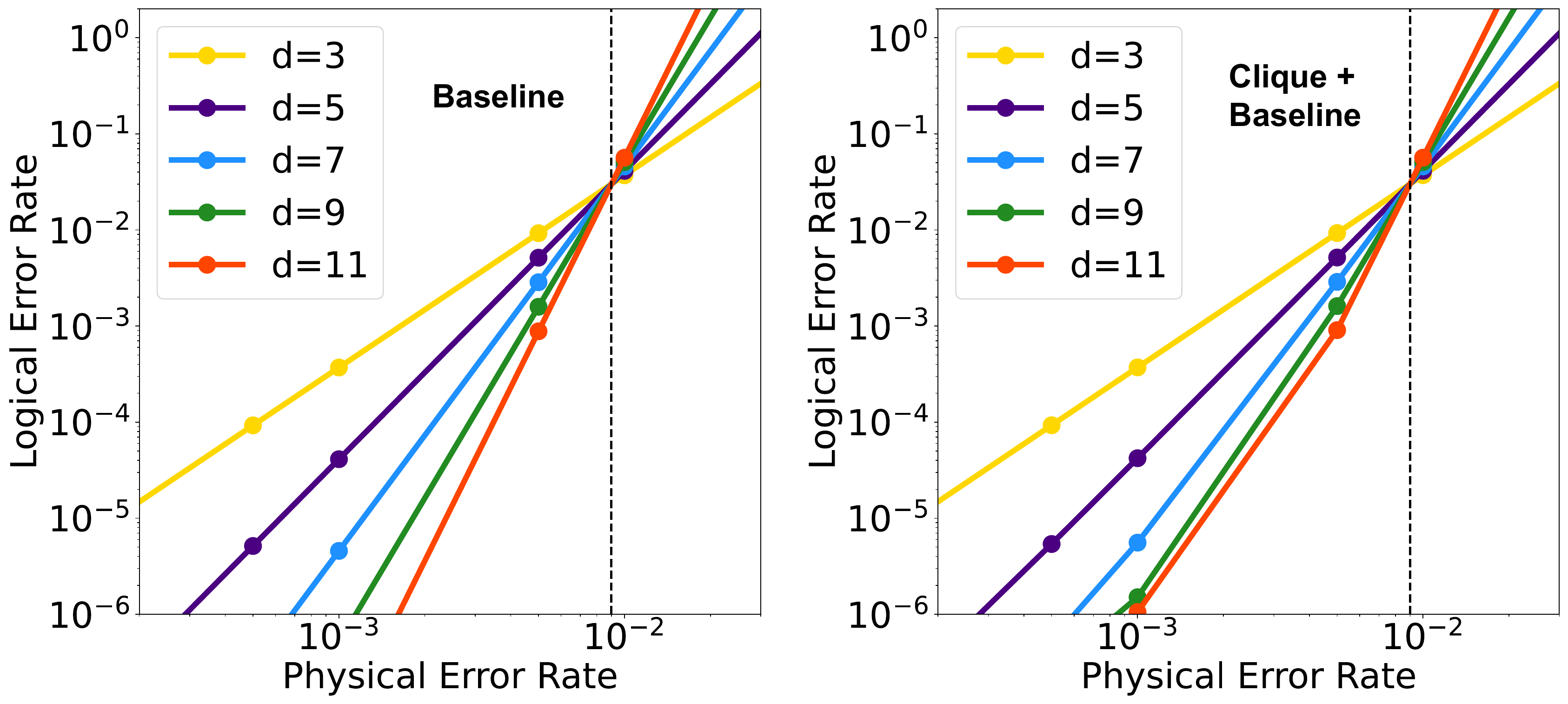}
\caption{
Logical Error Rate of Clique vs. baseline.
}
\label{fig:clique_ler}
\end{figure}

Next, we compare the accuracy of the Clique decoder to a traditional MWPM baseline in Fig.\ref{fig:clique_ler}.
Note that we assume that the our proposal uses the baseline decoder in the complex scenario while Clique is used in the trivial scenarios.
The complex vs. trivial decision is also made by Clique.
The comparison is shown for 5 different code distances, for a range of physical error rates and corresponding logical error rates.

Clearly, the Clique+Baseline setup performs almost exactly equivalent to the baseline for code distances d=3/5/7.
The accuracy of the Clique+Baseline is marginally worse compared to the baseline for higher code distances d=9/11.
\rev{The reason for the marginal worsening comes from Clique using only two rounds for measurement to achieve measurement error robustness. 
Therefore, measurement error which stick for greater than two cycles \emph{and} occur in a locally coordinated manner to look similar to a within-clique data qubit error, cause the Clique decoder to incorrectly assume this to be a trivial decode.
This is a fairly rare occurrence and thus Clique's deviation from the baseline is only evident for large code distances, since the MWPM baseline is tolerant to measurement errors that stick for up to `d' cycles.
If more rounds are used in Clique, further measurement error robustness can be achieved, enabling accuracy even closer to the baseline at high code distances.
Considering that the Clique hardware overheads are  30x lower than prior on-chip decoders (discussed next), additional measurement rounds can be added at limited cost, if required.}


\subsection{Decoder Overheads} 
\label{e:hw_cost}
\begin{figure}[t]
\centering
\includegraphics[width=0.99\columnwidth,trim={0.2cm 0cm 0.2cm 0cm},clip]{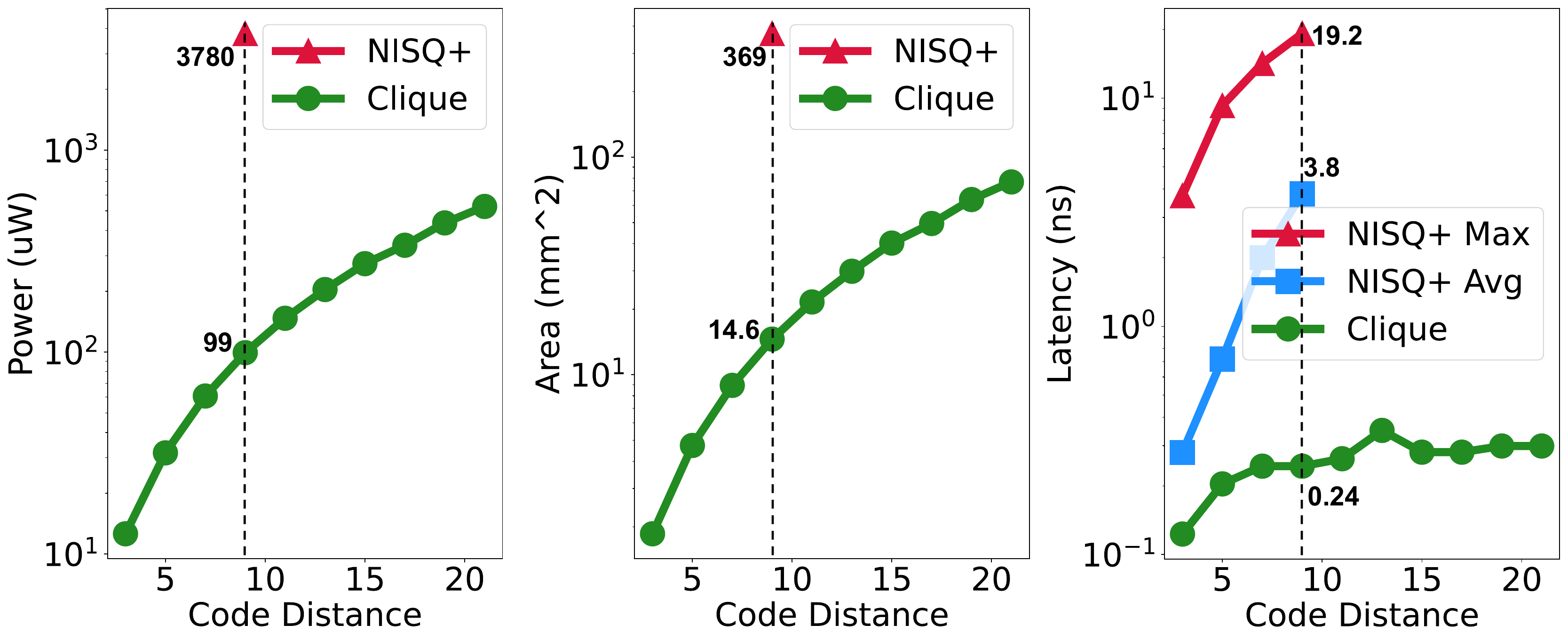}
\caption{
Power, Area and Latency overheads of Clique.
}
\label{fig:clique_sfq}
\end{figure}

Next, we evaluate power, area, and latency overheads for the SFQ implementation of Clique, for different code distances in Fig.\ref{fig:clique_sfq}. 
SFQ synthesis library details are provided in Table \ref{tab:SFQ_lib}.
We also compare against prior work NISQ+ with available data.

The power consumed per logical qubit for Clique varies from $10 \mu W$ to $500 \mu W$ for code distance ranging from 3 to 21.
Typical dilution refrigerators are capable of cooling up to 1 Watt of power at 4K temperature, so Clique should be able to support up to 2000 logical qubits at a code distance of 21 or 100,000 logical qubits at a code distance of 3.
Furthermore, at a code distance of 9, it is 37x more power efficient than the SFQ implementation of NISQ+.
Even though NISQ+ is an approximate decoder, it is provisioned to tackle worst case scenarios and is thus far more complicated than the simple Clique design.

In terms of area, Clique takes up under $100 mm^2$ per logical qubit even at a high code distance of 21.
Similar to power, it is 25x more area efficient than the NISQ+ design at a code distance of 9.
The Clique decoder only employs a few combinational logic gates, local to each clique, whereas NISQ+ requires more complicated communication between the many physical qubits in each logical qubit block.

Clique has a latency of 0.1 to 0.3 ns, and the latency is fairly fixed across all Clique decoding scenarios.
At a code distance of 9, the NISQ+ average latency is 15x higher than Clique.
Note that NISQ+ latency can be another 6x (multiplicative) worse in the worst-case decoding scenarios.

Thus, it is evident that Clique is a substantially more lightweight implementation compared to prior on-chip decoder implementations, due to its BTWC philosophy of tackling the trivial but very common decoding scenarios. 


\subsection{Bandwidth Allocation vs. Stalling Trade-offs}


\begin{figure}[t]
\centering
\includegraphics[width=0.99\columnwidth,trim={0cm 0cm 0cm 0cm},clip]{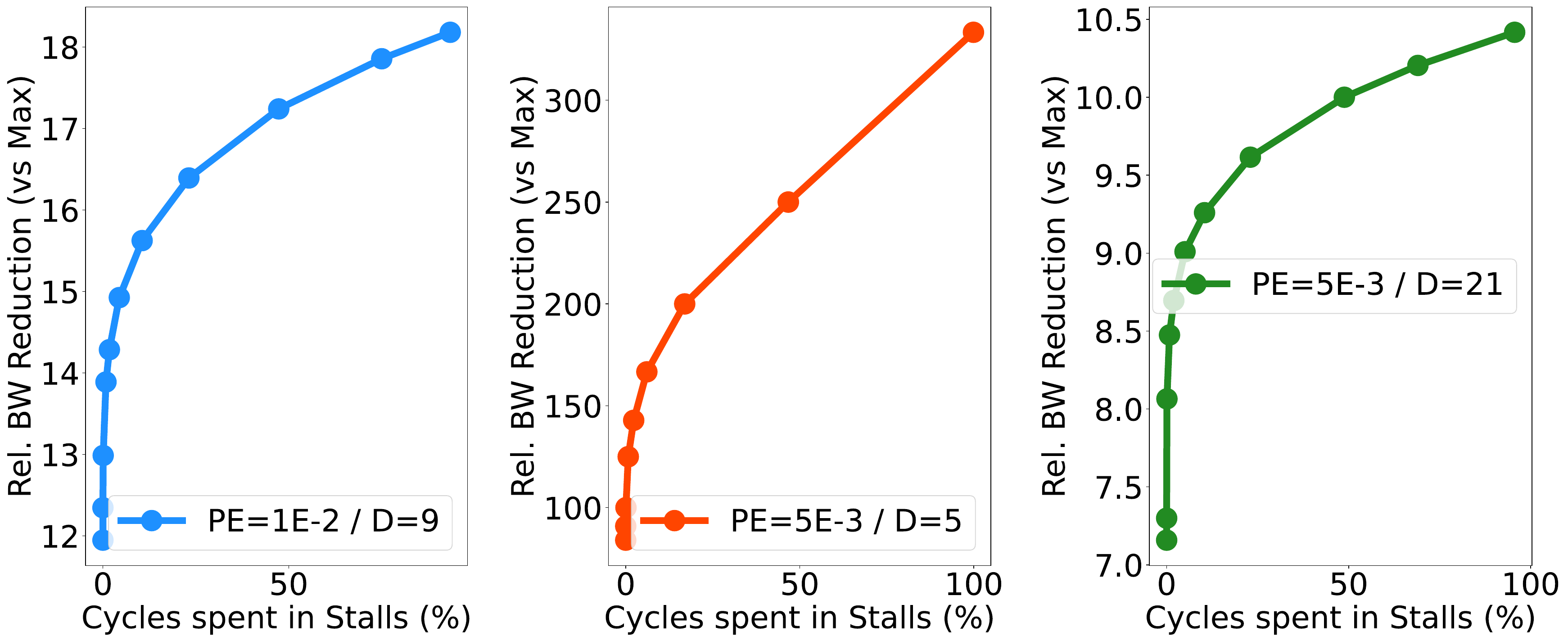}
\caption{
Bandwidth Allocation vs. Stalling Trade-offs
}
\label{fig:clique_bwstall}
\end{figure}

Finally, we evaluate the trade-off between off-chip bandwidth provisioning and execution stalling.
Fig.\ref{fig:clique_bwstall} shows the evaluation for three different combinations of physical error rate / code distance.
In all three instances, it is evident that provisioning for maximum bandwidth reduction (i.e., by setting the bandwidth strictly as per the average Clique decoding coverage) is not practical since this will lead to an infinite amount of stalling and the application will never complete execution due to the accumulating decode backlog.
This was discussed with an illustrative example in Section \ref{BWStall}.
Note that the increased execution time is not primarily a functional correctness issue since errors can be corrected, but a practical issue of the application execution never going to completion. 
This is, of course, not limited to Clique and is true for any optimization that improves the average case and not the worst case (including the syndrome compression in AFS~\cite{AFS}).

On the other hand, more conservative bandwidth provisioning is practical. 
For example, in the three instances shown, bandwidth reductions of 15x, 150x, and 8.5x can be achieved if the application / user is willing to tolerate an execution time increase of 10\% which is definitely practical, since the primary goal is fidelity and not performance.
These benefits are 1.5x, 2x, and 1.25x relatively lower than the maximum benefits(respectively), but can be achieved realistically.
It is also evident that the shape of the curve varies with physical error rates and code distance, so analyzing for the target system is important before making statistical trade-offs.
But at any point along the curve, the overwhelming benefits of the BTWC design are evident.

\section{Discussion}
\rev{The multi-level BTWC approach for QEC is applicable to any quantum computer that supports QEC. Today’s quantum devices do not support QEC (except for a few small demonstrations~\cite{GoogleQEC}) since the physical error rates need to be a bit lower and the device sizes have to be larger. But we expect that QEC based devices, with a few logical qubits and implemented with low code distances, are right around the corner.
Note that while the BTWC QEC philosophy is broadly applicable to all error codes and technologies, the specific Clique design we propose here targets surface codes, which are the QEC codes with the highest potential in the near/intermediate term, especially for superconducting transmon quantum devices.}

\subsection{Future Work}
While the Clique decoder specifically targets on-chip decoding for surface codes, BTWC opportunities can be explored beyond this scope:

\circled{1}\ The Clique decoder could be designed as an off-chip first-level decoder to handle the trivial common case decodes. This allows more flexibility in the implementation of the Clique decoder. While this is not beneficial from the bandwidth reduction perspective, it can help reduce decoding latency and/or improve energy efficiency. The reduced usage of the complex decoder would trivially reduce the power and energy requirements. Alternatively, the complex decoder could be run aggressively under looser power + thermal constraints, which would reduce decoding latency. Further, due to its reduced and more specialized usage, the complex decoder could be specifically designed for non-trivial scenarios. 

\circled{2}\ Especially in the off-chip scenario (but also for on-chip) there is opportunity to implement a deeper hierarchy of decoders with more specialization. Caching can also be explored in conjunction.

\circled{3}\ The Clique decoder is proposed to be implemented on-chip with classical hardware. Due to its trivial structure, it could potentially be implemented as a quantum circuit itself, using Toffoli gates. While this could have interesting implications, care should be taken since the decoder itself could then be prone to quantum errors. 

\circled{4}\ While this work targets surface codes, BTWC decoding could have suitability to other QEC codes. The Clique decoder specifically exploits sparsity and locality, both of which are high for surface codes at reasonable physical error rates. Other codes from the LDPC (Low-Density Parity-Check~\cite{Breuckmann_2021}) family often have sparsity but are not necessarily local (but provide other scalability benefits). However, suitable codes from the LDPC family for near-term QEC, might, in general, have to be designed with good locality due to the limited connectivity in quantum devices. This locality requirement could bode well in extending the Clique decoder to suit these LDPC codes.
The implementation specifics of the decoder will change with the structure of the code, but the wiring overheads could be kept under check if there is reasonable locality. In this regard, it should also be noted that the current Clique design has extremely low overheads --- some additional communication cost could easily be tolerated, especially as cryogenic technology continues to improve. Other codes such as Color codes~\cite{Bombin_2006}, especially for trapped-ion quantum technologies, are also worth exploring.

\subsection{Related Work}

Today's quantum devices are error-prone and up to around 100 qubits in size~\cite{preskill2018quantum}. 
Error mitigation strategies include, but are not limited to, noise-aware compilation~\cite{murali2019noise,tannu2019not}, scheduling for crosstalk~\cite{murali2020software,ding2020systematic}, 1Q gate scheduling in idle windows~\cite{smith2021error}, dynamical decoupling~\cite{viola1999dynamical,pokharel2018demonstration, souza2012robust}, zero-noise extrapolation~\cite{temme2017error,li2017efficient,giurgica2020digital}, readout error mitigation~\cite{tannu2019mitigating,bravyi2021mitigating}, exploiting quantum reversibility~\cite{patel2021qraft,smith2021error} and many more~\cite{czarnik2020error,Rosenberg2021,barron2020measurement,botelho2021error,wang2021error,takagi2021fundamental,temme2017error,li2017efficient,giurgica2020digital}.

State-of-the-art QEC decoding is achieved via Minimum Weight Perfect Matching (MWPM)~\cite{Dennis_2002} which is a graph pairing algorithm.
Recently, on-chip decoders have been proposed~\cite{NISQ+,QECOOL} to perform online decoding but suffer from challenging trade-offs of accuracy vs area/power budgets.
Off-chip decoders have been implemented via FPGAs~\cite{Lilliput}, LUTs~\cite{PhysRevA.90.062320}, special hardware designs~\cite{AFS} etc.

Prior work~\cite{Delfosse:2020} took first steps towards handling trivial error decodes with a `local' decoder that could also be used in a hierarchical setting. We differ from this work in multiple ways.
First, our work specifically proposes an on-chip hardware decoder which can be trivially implemented in the cryogenic domain with a simple array of gates.
\cite{Delfosse:2020} proposes a (decoding) graph reduction approach which iteratively runs over all the graph edges. 
As is, this would be implemented as a multi-step state machine, which is less suited to the cryogenic domain due to its state retention / memory requirements (to our knowledge, they target off-chip implementation).
This is an important distinction because our goal is specifically to reduce off-chip bandwidth---we do so by adding on-chip classical hardware, which is extremely lightweight and combinational, well suited to the cryogenic domain. 
Second, they do not handle measurement errors---our design can incorporate any number of measurement rounds, though we show evaluation results for two rounds.
Third, while they discuss decoding bandwidth and differentiate between the common-case and the worst-case, our proposal implements a practical end-to-end approach to handle the common, rare, and worst-case scenarios. 

Recently ~\cite{Paler:2022} proposed an off-chip lightweight pre-decoding step which re-weights the decoding graph according to likely correlations, and performs high confidence matching where possible.
Concurrent to our work, ~\cite{Smith:2022} propose a local `pre-decoder' based on cellular automata, which makes greedy corrections to reduce the amount of syndrome data sent to a standard matching decoder. 
In addition, ~\cite{Chamberland:2022} propose a neural network decoder that works as a local decoder that corrects errors arising from a constant number of faults, with longer error chains left to be corrected by a global decoder. 
Although clearly reducing the workload for the complex decoder in the off-chip setting, cryogenic on-chip hardware implementations of these proposals seem unclear but definitely worth abundant exploration.

While our work has focused on decoding for surface codes, real time decoding for other codes have been studied.
For example, software decoders have been used to achieve real-time decoding for color codes~\cite{Bombin_2006} on trapped-ion systems~\cite{arxiv.2107.07505} but have latencies less suited to superconducting devices.

\section{Conclusion}
The overheads of classical decoding for quantum error correction grow rapidly with the number of logical qubits and their correction code distance. Decoding at room temperature is bottle-necked by refrigerator I/O bandwidth while cryogenic on-chip decoding is bottle-necked  by power/thermal budget. 


We proposed {Better Than Worst-Case Decoding for Quantum Error Correction}, targeting Surface Codes, consisting of an On-chip Clique Decoder, Statistical Off-chip Bandwidth Allocation and Decode-Overflow Execution Stalling, which exploit trivial common QEC decode scenarios and achieve significant reduction in off-chip decode bandwidth at extremely low on-chip resource costs.

Importantly, we showcase that BTWC design is a critical step towards a practical scalable future for quantum error correction with tremendous potential beyond the specific scope of this paper targeting decoding for surface codes.

\section*{Acknowledgement}
This work is funded in part by EPiQC, an NSF Expedition in Computing, under award CCF-1730449; 
in part by STAQ under award NSF Phy-1818914; in part by NSF award 2110860; 
in part by the US Department of Energy Office  of Advanced Scientific Computing Research, Accelerated Research for Quantum Computing Program; 
and in part by the NSF Quantum Leap Challenge Institute for Hybrid Quantum Architectures and Networks (NSF Award 2016136) 
and in part based upon work supported by the U.S. Department of Energy, Office of Science, National Quantum Information Science Research Centers.  
This work was completed in part with resources provided by the University of Chicago’s Research Computing Center. 
GSR is supported as a Computing Innovation Fellow at the University of Chicago. This material is based upon work supported by the National Science Foundation under Grant \# 2030859 to the Computing Research Association for the CIFellows Project.
FTC is Chief Scientist for Quantum Software at ColdQuanta and an advisor to Quantum Circuits, Inc.

\bibliographystyle{plain}
\bibliography{references}

\begin{thebibliography}{10}

\bibitem{IBM_Blog2}
Expanding the ibm quantum roadmap to anticipate the future of quantum-centric
  supercomputing.
\newblock \url{https://research.ibm.com/blog/ibm-quantum-roadmap-2025}.

\bibitem{IBM_Blog1}
With fault tolerance the ultimate goal, error mitigation is the path that gets
  quantum computing to usefulness.
\newblock \url{https://research.ibm.com/blog/gammabar-for-quantum-advantage}.

\bibitem{GoogleQEC}
Rajeev Acharya, Igor Aleiner, Richard Allen, Trond~I. Andersen, Markus Ansmann,
  Frank Arute, Kunal Arya, Abraham Asfaw, Juan Atalaya, Ryan Babbush, Dave
  Bacon, Joseph~C. Bardin, Joao Basso, Andreas Bengtsson, Sergio Boixo, Gina
  Bortoli, Alexandre Bourassa, Jenna Bovaird, Leon Brill, Michael Broughton,
  Bob~B. Buckley, David~A. Buell, Tim Burger, Brian Burkett, Nicholas Bushnell,
  Yu~Chen, Zijun Chen, Ben Chiaro, Josh Cogan, Roberto Collins, Paul Conner,
  William Courtney, Alexander~L. Crook, Ben Curtin, Dripto~M. Debroy, Alexander
  Del~Toro Barba, Sean Demura, Andrew Dunsworth, Daniel Eppens, Catherine
  Erickson, Lara Faoro, Edward Farhi, Reza Fatemi, Leslie~Flores Burgos,
  Ebrahim Forati, Austin~G. Fowler, Brooks Foxen, William Giang, Craig Gidney,
  Dar Gilboa, Marissa Giustina, Alejandro~Grajales Dau, Jonathan~A. Gross,
  Steve Habegger, Michael~C. Hamilton, Matthew~P. Harrigan, Sean~D. Harrington,
  Oscar Higgott, Jeremy Hilton, Markus Hoffmann, Sabrina Hong, Trent Huang,
  Ashley Huff, William~J. Huggins, Lev~B. Ioffe, Sergei~V. Isakov, Justin
  Iveland, Evan Jeffrey, Zhang Jiang, Cody Jones, Pavol Juhas, Dvir Kafri,
  Kostyantyn Kechedzhi, Julian Kelly, Tanuj Khattar, Mostafa Khezri, Mária
  Kieferová, Seon Kim, Alexei Kitaev, Paul~V. Klimov, Andrey~R. Klots,
  Alexander~N. Korotkov, Fedor Kostritsa, John~Mark Kreikebaum, David Landhuis,
  Pavel Laptev, Kim-Ming Lau, Lily Laws, Joonho Lee, Kenny Lee, Brian~J.
  Lester, Alexander Lill, Wayne Liu, Aditya Locharla, Erik Lucero, Fionn~D.
  Malone, Jeffrey Marshall, Orion Martin, Jarrod~R. McClean, Trevor Mccourt,
  Matt McEwen, Anthony Megrant, Bernardo~Meurer Costa, Xiao Mi, Kevin~C. Miao,
  Masoud Mohseni, Shirin Montazeri, Alexis Morvan, Emily Mount, Wojciech
  Mruczkiewicz, Ofer Naaman, Matthew Neeley, Charles Neill, Ani Nersisyan,
  Hartmut Neven, Michael Newman, Jiun~How Ng, Anthony Nguyen, Murray Nguyen,
  Murphy~Yuezhen Niu, Thomas~E. O'Brien, Alex Opremcak, John Platt, Andre
  Petukhov, Rebecca Potter, Leonid~P. Pryadko, Chris Quintana, Pedram Roushan,
  Nicholas~C. Rubin, Negar Saei, Daniel Sank, Kannan Sankaragomathi, Kevin~J.
  Satzinger, Henry~F. Schurkus, Christopher Schuster, Michael~J. Shearn, Aaron
  Shorter, Vladimir Shvarts, Jindra Skruzny, Vadim Smelyanskiy, W.~Clarke
  Smith, George Sterling, Doug Strain, Marco Szalay, Alfredo Torres, Guifre
  Vidal, Benjamin Villalonga, Catherine~Vollgraff Heidweiller, Theodore White,
  Cheng Xing, Z.~Jamie Yao, Ping Yeh, Juhwan Yoo, Grayson Young, Adam Zalcman,
  Yaxing Zhang, and Ningfeng Zhu.
\newblock Suppressing quantum errors by scaling a surface code logical qubit,
  2022.

\bibitem{9906129}
Dorit Aharonov and Michael Ben-Or.
\newblock Fault-tolerant quantum computation with constant error rate, 1999.

\bibitem{austin2005opportunities}
Todd Austin, Valeria Bertacco, David Blaauw, and Trevor Mudge.
\newblock Opportunities and challenges for better than worst-case design.
\newblock In {\em Proceedings of the 2005 Asia and South Pacific Design
  Automation Conference}, pages 2--7, 2005.

\bibitem{barron2020measurement}
George~S. Barron and Christopher~J. Wood.
\newblock Measurement error mitigation for variational quantum algorithms,
  2020.

\bibitem{Bombin_2006}
H.~Bombin and M.~A. Martin-Delgado.
\newblock Topological quantum distillation.
\newblock {\em Physical Review Letters}, 97(18), oct 2006.

\bibitem{botelho2021error}
Ludmila Botelho, Adam Glos, Akash Kundu, Jarosław~Adam Miszczak, Özlem
  Salehi, and Zoltán Zimborás.
\newblock Error mitigation for variational quantum algorithms through
  mid-circuit measurements, 2021.

\bibitem{Surface-Codes}
S.~B. Bravyi and A.~Yu. Kitaev.
\newblock Quantum codes on a lattice with boundary, 1998.

\bibitem{bravyi2021mitigating}
Sergey Bravyi, Sarah Sheldon, Abhinav Kandala, David~C Mckay, and Jay~M
  Gambetta.
\newblock Mitigating measurement errors in multiqubit experiments.
\newblock {\em Physical Review A}, 103(4):042605, 2021.

\bibitem{Breuckmann_2021}
Nikolas~P. Breuckmann and Jens~Niklas Eberhardt.
\newblock Quantum low-density parity-check codes.
\newblock {\em {PRX} Quantum}, 2(4), oct 2021.

\bibitem{XQSim}
Ilkwon Byun, Junpyo Kim, Dongmoon Min, Ikki Nagaoka, Kosuke Fukumitsu, Iori
  Ishikawa, Teruo Tanimoto, Masamitsu Tanaka, Koji Inoue, and Jangwoo Kim.
\newblock Xqsim: Modeling cross-technology control processors for 10+k qubit
  quantum computers.
\newblock In {\em Proceedings of the 49th Annual International Symposium on
  Computer Architecture}, ISCA '22, page 366–382, New York, NY, USA, 2022.
  Association for Computing Machinery.

\bibitem{Chamberland:2022}
Christopher Chamberland, Luis Goncalves, Prasahnt Sivarajah, Eric Peterson, and
  Sebastian Grimberg.
\newblock Techniques for combining fast local decoders with global decoders
  under circuit-level noise, 2022.

\bibitem{cong2014better}
Jason Cong, Henry Duwe, Rakesh Kumar, and Sen Li.
\newblock Better-than-worst-case design: Progress and opportunities.
\newblock {\em Journal of computer science and technology}, 29(4):656--663,
  2014.

\bibitem{czarnik2020error}
Piotr Czarnik, Andrew Arrasmith, Patrick~J. Coles, and Lukasz Cincio.
\newblock Error mitigation with clifford quantum-circuit data, 2020.

\bibitem{Lilliput}
Poulami Das, Aditya Locharla, and Cody Jones.
\newblock Lilliput: A lightweight low-latency lookup-table based decoder for
  near-term quantum error correction, 2021.

\bibitem{AFS}
Poulami Das, Christopher~A. Pattison, Srilatha Manne, Douglas~M. Carmean,
  Krysta~M. Svore, Moinuddin Qureshi, and Nicolas Delfosse.
\newblock Afs: Accurate, fast, and scalable error-decoding for fault-tolerant
  quantum computers.
\newblock In {\em 2022 IEEE International Symposium on High-Performance
  Computer Architecture (HPCA)}, pages 259--273, 2022.

\bibitem{Delfosse:2020}
Nicolas Delfosse.
\newblock Hierarchical decoding to reduce hardware requirements for quantum
  computing, 2020.

\bibitem{Dennis_2002}
Eric Dennis, Alexei Kitaev, Andrew Landahl, and John Preskill.
\newblock Topological quantum memory.
\newblock {\em Journal of Mathematical Physics}, 43(9):4452--4505, sep 2002.

\bibitem{ding2020quantum}
Yongshan Ding and Frederic~T Chong.
\newblock Quantum computer systems: Research for noisy intermediate-scale
  quantum computers.
\newblock {\em Synthesis Lectures on Computer Architecture}, 15(2):1--227,
  2020.

\bibitem{ding2020systematic}
Yongshan Ding, Pranav Gokhale, Sophia~Fuhui Lin, Richard Rines, Thomas Propson,
  and Frederic~T Chong.
\newblock Systematic crosstalk mitigation for superconducting qubits via
  frequency-aware compilation.
\newblock {\em arXiv preprint arXiv:2008.09503}, 2020.

\bibitem{VLQ}
Casey Duckering, Jonathan~M. Baker, David~I. Schuster, and Frederic~T. Chong.
\newblock Virtualized logical qubits: A 2.5d architecture for error-corrected
  quantum computing, 2020.

\bibitem{farhi2014quantum}
Edward Farhi, Jeffrey Goldstone, and Sam Gutmann.
\newblock A quantum approximate optimization algorithm, 2014.

\bibitem{Fowler_2012}
Austin~G. Fowler, Matteo Mariantoni, John~M. Martinis, and Andrew~N. Cleland.
\newblock Surface codes: Towards practical large-scale quantum computation.
\newblock {\em Physical Review A}, 86(3), sep 2012.

\bibitem{giurgica2020digital}
Tudor Giurgica-Tiron, Yousef Hindy, Ryan LaRose, Andrea Mari, and William~J
  Zeng.
\newblock Digital zero noise extrapolation for quantum error mitigation.
\newblock In {\em 2020 IEEE International Conference on Quantum Computing and
  Engineering (QCE)}, pages 306--316. IEEE, 2020.

\bibitem{Grover96afast}
Lov~K. Grover.
\newblock A fast quantum mechanical algorithm for database search.
\newblock In {\em ANNUAL ACM SYMPOSIUM ON THEORY OF COMPUTING}, pages 212--219.
  ACM, 1996.

\bibitem{NISQ+}
Adam Holmes, Mohammad~Reza Jokar, Ghasem Pasandi, Yongshan Ding, Massoud
  Pedram, and Frederic~T. Chong.
\newblock Nisq+: Boosting quantum computing power by approximating quantum
  error correction, 2020.

\bibitem{ERSFQ}
Alex Kirichenko, Saad Sarwana, and Igor Vernik.
\newblock Ersfq-zero static power dissipation single flux quantum logic.
\newblock 03 2012.

\bibitem{Krinner_2019}
S.~Krinner, S.~Storz, P.~Kurpiers, P.~Magnard, J.~Heinsoo, R.~Keller,
  J.~Lütolf, C.~Eichler, and A.~Wallraff.
\newblock Engineering cryogenic setups for 100-qubit scale superconducting
  circuit systems.
\newblock {\em {EPJ} Quantum Technology}, 6(1), may 2019.

\bibitem{li2017efficient}
Ying Li and Simon~C. Benjamin.
\newblock Efficient variational quantum simulator incorporating active error
  minimization.
\newblock {\em Phys. Rev. X}, 7:021050, Jun 2017.

\bibitem{murali2019noise}
Prakash Murali, Jonathan~M Baker, Ali Javadi-Abhari, Frederic~T Chong, and
  Margaret Martonosi.
\newblock Noise-adaptive compiler mappings for noisy intermediate-scale quantum
  computers.
\newblock In {\em Proceedings of the Twenty-Fourth International Conference on
  Architectural Support for Programming Languages and Operating Systems}, pages
  1015--1029, 2019.

\bibitem{murali2020software}
Prakash Murali, David~C McKay, Margaret Martonosi, and Ali Javadi-Abhari.
\newblock Software mitigation of crosstalk on noisy intermediate-scale quantum
  computers.
\newblock In {\em Proceedings of the Twenty-Fifth International Conference on
  Architectural Support for Programming Languages and Operating Systems}, pages
  1001--1016, 2020.

\bibitem{mike_ike_2020}
Michael~A Nielsen and Isaac Chuang.
\newblock {\em Quantum computation and quantum information}.
\newblock Cambridge University Press, 2010.

\bibitem{O_Gorman_2017}
Joe O’Gorman and Earl~T. Campbell.
\newblock Quantum computation with realistic magic-state factories.
\newblock {\em Physical Review A}, 95(3), Mar 2017.

\bibitem{Paler:2022}
Alexandru Paler and Austin~G. Fowler.
\newblock Pipelined correlated minimum weight perfect matching of the surface
  code, 2022.

\bibitem{8351603}
Ghasem Pasandi, Alireza Shafaei, and Massoud Pedram.
\newblock Sfqmap: A technology mapping tool for single flux quantum logic
  circuits.
\newblock In {\em 2018 IEEE International Symposium on Circuits and Systems
  (ISCAS)}, pages 1--5, 2018.

\bibitem{patel2021qraft}
Tirthak Patel and Devesh Tiwari.
\newblock Qraft: reverse your quantum circuit and know the correct program
  output.
\newblock In {\em Proceedings of the 26th ACM International Conference on
  Architectural Support for Programming Languages and Operating Systems}, pages
  443--455, 2021.

\bibitem{peruzzo2014variational}
Alberto Peruzzo, Jarrod McClean, Peter Shadbolt, Man-Hong Yung, Xiao-Qi Zhou,
  Peter~J Love, Al{\'a}n Aspuru-Guzik, and Jeremy~L O’brien.
\newblock A variational eigenvalue solver on a photonic quantum processor.
\newblock {\em Nature communications}, 5:4213, 2014.

\bibitem{pokharel2018demonstration}
Bibek Pokharel, Namit Anand, Benjamin Fortman, and Daniel~A Lidar.
\newblock Demonstration of fidelity improvement using dynamical decoupling with
  superconducting qubits.
\newblock {\em Physical review letters}, 121(22):220502, 2018.

\bibitem{preskill2018quantum}
John Preskill.
\newblock Quantum computing in the nisq era and beyond.
\newblock {\em Quantum}, 2:79, 2018.

\bibitem{Rosenberg2021}
Eliott Rosenberg, Paul Ginsparg, and Peter~L. McMahon.
\newblock Experimental error mitigation using linear rescaling for variational
  quantum eigensolving with up to 20 qubits.
\newblock {\em Quantum Science and Technology}, Nov 2021.

\bibitem{arxiv.2107.07505}
C.~Ryan-Anderson, J.~G. Bohnet, K.~Lee, D.~Gresh, A.~Hankin, J.~P. Gaebler,
  D.~Francois, A.~Chernoguzov, D.~Lucchetti, N.~C. Brown, T.~M. Gatterman,
  S.~K. Halit, K.~Gilmore, J.~Gerber, B.~Neyenhuis, D.~Hayes, and R.~P. Stutz.
\newblock Realization of real-time fault-tolerant quantum error correction,
  2021.

\bibitem{Shor_1997}
Peter~W. Shor.
\newblock Polynomial-time algorithms for prime factorization and discrete
  logarithms on a quantum computer.
\newblock {\em SIAM Journal on Computing}, 26(5):1484–1509, Oct 1997.

\bibitem{smith2021error}
Kaitlin~N Smith, Gokul~Subramanian Ravi, Prakash Murali, Jonathan~M Baker,
  Nathan Earnest, Ali Javadi-Abhari, and Frederic~T Chong.
\newblock Error mitigation in quantum computers through instruction scheduling.
\newblock {\em arXiv preprint arXiv:2105.01760}, 2021.

\bibitem{Smith:2022}
Samuel~C. Smith, Benjamin~J. Brown, and Stephen~D. Bartlett.
\newblock A local pre-decoder to reduce the bandwidth and latency of quantum
  error correction, 2022.

\bibitem{souza2012robust}
Alexandre~M Souza, Gonzalo~A {\'A}lvarez, and Dieter Suter.
\newblock Robust dynamical decoupling.
\newblock {\em Philosophical Transactions of the Royal Society A: Mathematical,
  Physical and Engineering Sciences}, 370(1976):4748--4769, 2012.

\bibitem{takagi2021fundamental}
Ryuji Takagi, Suguru Endo, Shintaro Minagawa, and Mile Gu.
\newblock Fundamental limits of quantum error mitigation, 2021.

\bibitem{tannu2019mitigating}
Swamit~S Tannu and Moinuddin~K Qureshi.
\newblock Mitigating measurement errors in quantum computers by exploiting
  state-dependent bias.
\newblock In {\em Proceedings of the 52nd Annual IEEE/ACM International
  Symposium on Microarchitecture}, pages 279--290, 2019.

\bibitem{tannu2019not}
Swamit~S Tannu and Moinuddin~K Qureshi.
\newblock Not all qubits are created equal: a case for variability-aware
  policies for nisq-era quantum computers.
\newblock In {\em Proceedings of the Twenty-Fourth International Conference on
  Architectural Support for Programming Languages and Operating Systems}, pages
  987--999, 2019.

\bibitem{temme2017error}
Kristan Temme, Sergey Bravyi, and Jay~M Gambetta.
\newblock Error mitigation for short-depth quantum circuits.
\newblock {\em Physical review letters}, 119(18):180509, 2017.

\bibitem{PhysRevA.90.062320}
Yu~Tomita and Krysta~M. Svore.
\newblock Low-distance surface codes under realistic quantum noise.
\newblock {\em Phys. Rev. A}, 90:062320, Dec 2014.

\bibitem{QECOOL}
Yosuke Ueno, Masaaki Kondo, Masamitsu Tanaka, Yasunari Suzuki, and Yutaka
  Tabuchi.
\newblock Qecool: On-line quantum error correction with a superconducting
  decoder for surface code.
\newblock In {\em 2021 58th ACM/IEEE Design Automation Conference (DAC)}, pages
  451--456, 2021.

\bibitem{qulatis}
Yosuke Ueno, Masaaki Kondo, Masamitsu Tanaka, Yasunari Suzuki, and Yutaka
  Tabuchi.
\newblock Qulatis: A quantum error correction methodology toward lattice
  surgery.
\newblock In {\em 2022 IEEE International Symposium on High-Performance
  Computer Architecture (HPCA)}, pages 274--287, 2022.

\bibitem{viola1999dynamical}
Lorenza Viola, Emanuel Knill, and Seth Lloyd.
\newblock Dynamical decoupling of open quantum systems.
\newblock {\em Physical Review Letters}, 82(12):2417, 1999.

\bibitem{wang2021error}
Samson Wang, Piotr Czarnik, Andrew Arrasmith, M.~Cerezo, Lukasz Cincio, and
  Patrick~J. Coles.
\newblock Can error mitigation improve trainability of noisy variational
  quantum algorithms?, 2021.

\end{thebibliography}

\end{document}